\documentclass[journal,10pt,twocolumn]{IEEEtran}
\usepackage{amssymb}
\pdfoutput=1
\usepackage{algorithm}
\usepackage{algpseudocode}
\usepackage{algorithmicx}
\usepackage{amsfonts}
\usepackage{amsmath}
\usepackage{amssymb}
\usepackage{amsthm}
\usepackage{array}
\usepackage{bbding}
\usepackage{bigints}
\usepackage{booktabs}
\usepackage{cite}
\usepackage{cleveref}
\usepackage{color}
\usepackage{diagbox}
\usepackage{epsfig,latexsym}
\usepackage{epstopdf}
\usepackage{graphicx}
\usepackage{fancyhdr}
\usepackage{float}
\usepackage{flushend}
\usepackage{indentfirst}
\usepackage{lastpage}
\usepackage{makecell}
\usepackage{mathtools}
\usepackage{multirow}
\usepackage{pifont}
\usepackage{psfrag}
\usepackage{setspace}
\usepackage{stfloats}
\usepackage{subfloat}
\usepackage{times}
\usepackage{subfigure}

\newcommand{\argmin}{\arg\,\min}
\newcommand{\argmax}{\arg\,\max}

\newcommand{\algorithmicbreak}{\textbf{break}}
\newcommand{\BREAK}{\State \algorithmicbreak}

\begin{document}
\title{Space-Time Shift Keying Aided OTFS Modulation for Orthogonal Multiple Access}

\author{{Zeping Sui}, {\em Member,~IEEE}, {Hongming~Zhang}, {\em Senior Member,~IEEE}, {Sumei Sun}, {\em Fellow,~IEEE}, {Lie-Liang Yang}, {\em Fellow,~IEEE}, and {Lajos Hanzo}, {\em Life Fellow,~IEEE}
\thanks{Zeping Sui is with the Institute of Acoustics, Chinese Academy of Sciences, Beijing 100190, China; and also with the University of Chinese Academy of Sciences, Beijing 100049, China (e-mail: suizeping@mail.ioa.ac.cn).}
\thanks{Hongming Zhang is with the School of Information and Communication Engineering, Beijing University of Posts and Telecommunications, Beijing 100876, China (email: zhanghm@bupt.edu.cn).}
\thanks{Sumei Sun is with the Institute for Infocomm Research, Agency for Science, Technology and Research, Singapore 138632 (e-mail: sunsm@i2r.a-star.edu.sg).}
\thanks{Lie-Liang Yang and Lajos Hanzo are with the Department of Electronics and Computer Science, University of Southampton, Southampton SO17 1BJ, U.K. (e-mail: lly@ecs.soton.ac.uk; lh@ecs.soton.ac.uk).}%
}
\maketitle

\begin{abstract}
Space-time shift keying-aided orthogonal time frequency space modulation-based multiple access (STSK-OTFS-MA) is proposed for reliable uplink transmission in high-Doppler scenarios. As a beneficial feature of our STSK-OTFS-MA system, extra information bits are mapped onto the indices of the active dispersion matrices, which allows the system to enjoy the joint benefits of both STSK and OTFS signalling. Due to the fact that both the time-, space- and DD-domain degrees of freedom are jointly exploited, our STSK-OTFS-MA achieves increased diversity and coding gains. To mitigate the potentially excessive detection complexity, the sparse structure of the equivalent transmitted symbol vector is exploited, resulting in a pair of low-complexity near-maximum likelihood (ML) multiuser detection algorithms. Explicitly, we conceive a progressive residual check-based greedy detector (PRCGD) and an iterative reduced-space check-based detector (IRCD). Then, we derive both the unconditional single-user pairwise error probability (SU-UPEP) and a tight bit error ratio (BER) union-bound for our single-user STSK-OTFS-MA system employing the ML detector. Furthermore, the discrete-input continuous-output memoryless channel (DCMC) capacity of the proposed system is derived. The optimal dispersion matrices (DMs) are designed based on the maximum attainable diversity and coding gain metrics. Finally, it is demonstrated that our STSK-OTFS-MA system achieves both a lower BER and a higher DCMC capacity than its conventional spatial modulation (SM) {and its orthogonal frequency-division multiplexing (OFDM) counterparts. As a benefit, the proposed system strikes a compelling BER \emph{vs.} system complexity as well as BER \emph{vs.} detection complexity trade-offs.}
\end{abstract}
\IEEEpeerreviewmaketitle

\section{Introduction}\label{Section1}
During the last decade, space-time shift keying (STSK) \cite{5599264,6316188,elfadil2018trellis,8322306} has been considered a compelling multi-functional multiple-input multiple-output (MIMO) arrangement, where the information bits are jointly mapped onto the conventional amplitude-phase modulated (APM) symbols and the indices of the active dispersion matrices (DMs). To elaborate further, each single APM symbol is dispersed to multiple transmit antennas (TAs) and time-slots by activating one out of $Q$ DMs. Hence, STSK can achieve diversity and multiplexing gains \cite{5599264}. By contrast, the conventional spatial modulation (SM) activates one TA to transmit a single APM symbol, yielding only receive diversity gain \cite{4382913}.

Orthogonal time frequency space (OTFS) modulation also constitutes a promising candidate for next-generation wireless networks \cite{7925924,8424569,9404861}. Since it is capable of providing reliable transmission in high-mobility scenarios, it has been widely studied in the context of reconfigurable intelligent surfaces (RISs) \cite{9864300} and low-earth orbit (LEO) satellites \cite{9689960}. More specifically, in OTFS systems, the information symbols are mapped to the delay-Doppler (DD)-domain, and each symbol is spread across the entire time-frequency (TF)-domain by leveraging the inverse symplectic finite Fourier transform (ISFFT). Therefore, OTFS can attain both time and frequency diversity gains, if channels are time-frequency selective or doubly-selective \cite{9508932,9082873}. In addition, the sparse DD-domain sparse representations of the doubly-selective channels are incorporated into the OTFS theory, and the dimension of the DD-domain channel model is reduced to the number of resolvable paths \cite{7925924,8424569}. Since the communication distance and relative velocity can be approximated as constants within a few milliseconds, the DD-domain channels can be regarded as nearly time-invariant over an entire OTFS frame \cite{9492800}. Furthermore, the inter-carrier interference (ICI) and inter-symbol interference (ISI) introduced by Doppler and delay spreads remain quasi-orthogonal with the aid of ISFFT, which can hence be processed in the Doppler-domain and delay-domain separately \cite{7925924}. By contrast, the performance of the conventional orthogonal frequency-division multiplexing (OFDM) suffers severely in the face of high-Doppler doubly-selective channels, since the orthogonality of subcarriers may be destroyed by severe ICI. Therefore, in high-Doppler scenarios, OTFS constitutes a more promising signaling scheme than the conventional OFDM \cite{7925924,8424569}.

More recently, OTFS-based non-orthogonal multiple-access (NOMA) communication schemes have been studied \cite{9411900,9794710,8786203,10022044}, where different users arranged in the same DD-domain resource blocks (RBs) are distinguished by their unique sparse codewords \cite{9411900,9794710,10022044} and power levels \cite{8786203}. However, the performance of OTFS-NOMA may degrade significantly due to the non-orthogonality-induced interference, and the excessive complexity of the transceiver \cite{9860075}. As a remedy, OTFS-based orthogonal MA (OTFS-OMA) techniques have been designed in \cite{8515088,9860075}, where the RBs of different users are arranged in non-overlapping grids in the TF- and/or DD-domains. In \cite{9860075}, the spectral efficiency (SE) of several OTFS-OMA schemes using rectangular pulse shapes are analyzed. Nevertheless, the above OTFS-OMA schemes only modulate data in the DD-domain. None of them exploits the degrees of freedom in the space-time (ST)-domain, even though it would be beneficial to leverage both transmit and receive diversity gains for further BER performance improvement.

As a parallel development, MA communications schemes have been designed in association with SM/STSK in \cite{6316188,7448828}. Specifically in \cite{6316188}, the STSK-aided OFDM-based multiple access (STSK-OFDM-MA) paradigms have been proposed, where an attractive diversity \emph{vs.} multiplexing gain trade-off was provided. However, these SM/STSK-based schemes are designed based on flat Rayleigh fading or \emph{low-mobility} frequency-selective fading channels, but without considering the ICI imposed by \emph{high-mobility environments}. {However, next-generation MA systems aim for providing reliable data transmission under high-mobility scenarios \cite{9496190,10043628}. Therefore, the BER performance of the STSK-OFDM-MA schemes may degrade substantially under doubly-selective channels, yielding a significant system reliability loss. On the other hand, it can be observed from \cite{6316188} that the STSK-OFDM-MA scheme is capable of providing better BER performance than the conventional solutions, resulting in a more reliable communication system. This observation implies that the reliability of the above-mentioned OTFS-OMA systems can be further enhanced by exploiting the STSK technique to attain both higher diversity and coding gains.} Against this backdrop, in this paper we jointly invoke the DD-, space- and time-domain (TD) resources for transmission over doubly-selective channels. Explicitly, by intrinsically amalgamating STSK and OTFS-OMA, we propose space-time shift keying-aided OTFS-based MA (STSK-OTFS-MA) for reliable communications in doubly-selective channels.

The novel contributions of the paper are boldly and explicitly contrasted to the existing literatures in Table \ref{table1}, which are addressed below.
\begin{table*}[t]
\footnotesize
\centering
\caption{Contrasting our contributions to the literature}
\label{table1}
\begin{tabular}{l|c|c|c|c|c|c|c|c}
\hline
Contributions & \textbf{This paper} & \cite{6316188} & \cite{8424569} & \cite{9404861} & \cite{9860075} & \cite{8515088} & \cite{7448828} & \cite{8686339} \\
\hline
\hline
OTFS-OMA & \checkmark &  &  &  & \checkmark & \checkmark &  & \\  
\hline
STSK-aided MA & \checkmark & \checkmark &  &  &  &  & \checkmark & \\  
\hline
BER performance analysis & \checkmark &  &  & \checkmark &  & &  & \checkmark\\  
\hline
Coding gain \& diversity order analysis & \checkmark &  & & \checkmark &  & & \checkmark & \checkmark\\  
\hline
Greedy check detection & \checkmark &  &   &  &  & &  & \\  
\hline 
Reduced-space check detection & \checkmark &  & &  &  & & & \\  
\hline
Capacity analysis & \checkmark &  &  &  &  & &  & \\ 
\hline
Complexity analysis & \checkmark &  & \checkmark &  &  & &  & \\ 
\hline
Dispersion matrix design & \checkmark &  &  &  &  & & &    \\ 
\hline
LDPC coded system & \checkmark & \checkmark &  & \checkmark &  & & &    \\ 
\hline
\end{tabular}
\end{table*}
{\begin{itemize}
\item We propose an STSK-OTFS-MA scheme for supporting the reliable data transmission of multiple users over high-mobility channels, where information is conveyed both by the classic APM symbols and the indices of active DMs. According to the $(N\times M)$ DD-domain grids, we first activate one out of $Q$ DMs for spreading $NM$ APM symbols to both the space and time dimensions, resulting in $NM$ STSK blocks. Then a tailor-made ST mapper is conceived for mapping the elements of the STSK blocks onto the transmitted DD-domain symbol matrices of users, which enables our STSK-OTFS-MA to achieve both transmit and receive diversity gains. Moreover, based on the DD-domain statistics of users, a resource-allocation scheme is introduced for the STSK-OTFS-MA system. Explicitly, the RBs of different users are mapped to the non-overlapping grids along the delay domain to mitigate the multiuser interference (MUI) caused by the Doppler shift. Furthermore, the proposed STSK-OTFS-MA is capable of striking an attractive diversity versus multiplexing gain trade-off. Both the analytical and simulation results illustrated that the STSK-OTFS-MA advocated achieves a better BER performance than its conventional SM-OTFS, single-input-multiple-output (SIMO)-OTFS and {STSK-OFDM-MA} counterparts. {Additionally, the general flexibility of the proposed STSK-OTFS-MA scheme is demonstrated in different-rate low-density parity-check (LDPC)-coded systems. Finally, the BER \emph{vs.} system complexity of the STSK-OTFS-MA and other counterparts are also evaluated.}
\item A pair of low-complexity near-maximum likelihood detectors (MLDs) are proposed for the STSK-OTFS-MA scheme. Firstly, inspired by the family of greedy algorithms designed for compressed sensing (CS), a progressive residual check-based greedy detector (PRCGD) is conceived, where optimal local choices are obtained at each iteration, yielding a detector approaching the globally optimal performance. Furthermore, commencing with the consideration of detecting the APM and DM-index symbols separately, an iterative reduced-space check-based detector (IRCD) is proposed. Specifically, by sorting the reliabilities of all DM activation patterns (DAPs), a reduced set of DAPs is tested. Finally, the BER performance \emph{vs.} complexity of the MLD, of the PRCGD and of the IRCD are compared.
\item We derive the unconditional single-user pairwise error probability (SU-UPEP) of the STSK-OTFS-MA system. Then by invoking the union-bound technique, we derive the closed-form single-user BER bound of our proposed STSK-OTFS-MA system employing MLD, which is shown to be tight for moderate to high signal-to-noise ratios (SNRs). Then, based on the SU-UPEP, both the diversity order and the ST coding gain achieved by the STSK-OTFS-MA system are determined.
\item The single-user discrete-input continuous-output memoryless channel (DCMC) capacity of the STSK-OTFS-MA system is derived, which is demonstrated to outperform its SM counterpart. Additionally, based on the SU-UPEP and the DCMC capacity derived, the design criteria of DMs are proposed, for approaching the maximum attainable diversity order and ST coding gain.
\end{itemize}}

The rest of the paper is structured as follows. In Section \ref{Section2}, the proposed STSK-OTFS-MA system model is investigated. Then, low-complexity near-ML multiuser detection algorithms are proposed in Section \ref{Section3}. In Section \ref{Section4}, the overall system performance is characterized and the DM design algorithm is detailed. The simulation results are shown in Section \ref{Section5}. Finally, the conclusions are offered in Section \ref{Section6}.

\emph{Notation:} We use the following notation throughout this paper: $\mathbb{C}$ and $\mathbb{R}$ are the ring of complex and real; $\mathbb{B}$ and $\mathbb{Z}^{M}_{+}$ represent the real integer set of $\{1,\ldots,M\}$ and the bit set consisting of $\left\{0,1\right\}$; $\mathbb{E}[\cdot]$ and $\text{tr}\left\{\cdot\right\}$ denote the expectation and trace operator; $x(l)$ and $X(l,k)$ are the $l$th of vector $\pmb{x}$ and $(l,k)$th element of matrix $\pmb{X}$, respectively; $\text{vec}(\pmb{A})$ denotes the vector formulated by stacking the columns of $\pmb{A}$ to obtain a single column vector matrix, and $\text{vec}^{-1}(\pmb{a})$ denotes the inverse vectorization operation to form the original matrix; $\otimes$ denotes the Kronecker product of two matrices; $\mathcal{CN}(\pmb{a},\pmb{B})$ is the complex Gaussian distribution having a mean vector $\pmb{a}$ and covariance matrix $\pmb{B}$; $\pmb{A}[:,1:n]$ and $\pmb{A}[1:m,:]$ represent the first $n$ columns and first $m$ rows of a matrix $\pmb{A}$, respectively; $\pmb{I}_N$ and $\pmb{I}_N(l)$ denote an $N$-dimensional identity matrix and its rows shift by $l$; The module-$N$ and determinant operations are defined by $\left[\cdot\right]_N$ and $\text{det}(\cdot)$; $\delta(\cdot)$ is the delta function; The uniform distribution in the interval $[a,b]$ is denoted by $\mathcal{U}[a,b]$; $\text{sta}\{\pmb{A}_n^{(u)}\}|_{n=0}^{N-1}=[(\pmb{A}_0^{(u)})^T,(\pmb{A}_1^{(u)})^T,\ldots,(\pmb{A}_{N-1}^{(u)})^T]^T$ and $\text{sta}\{\pmb{A}_n^{(u)}\}|_{u=0}^{U-1}=[(\pmb{A}_n^{(0)})^T,(\pmb{A}_n^{(1)})^T,\ldots,(\pmb{A}_n^{(U-1)})^T]^T$ denotes the matrix (vector) formulated by stacking $N$ and $U$ identical-dimensional sub-matrices (sub-vectors) $\pmb{A}_n$ and $\pmb{A}_n^{(u)}$ for $n=0,\ldots,N-1$ and $u=0,\ldots,N-1$, respectively. 
\section{System Model}\label{Section2}
\subsection{Transmitter Description}\label{Section2-1}	
\begin{figure*}[t]
\centering
\includegraphics[width=0.8\linewidth]{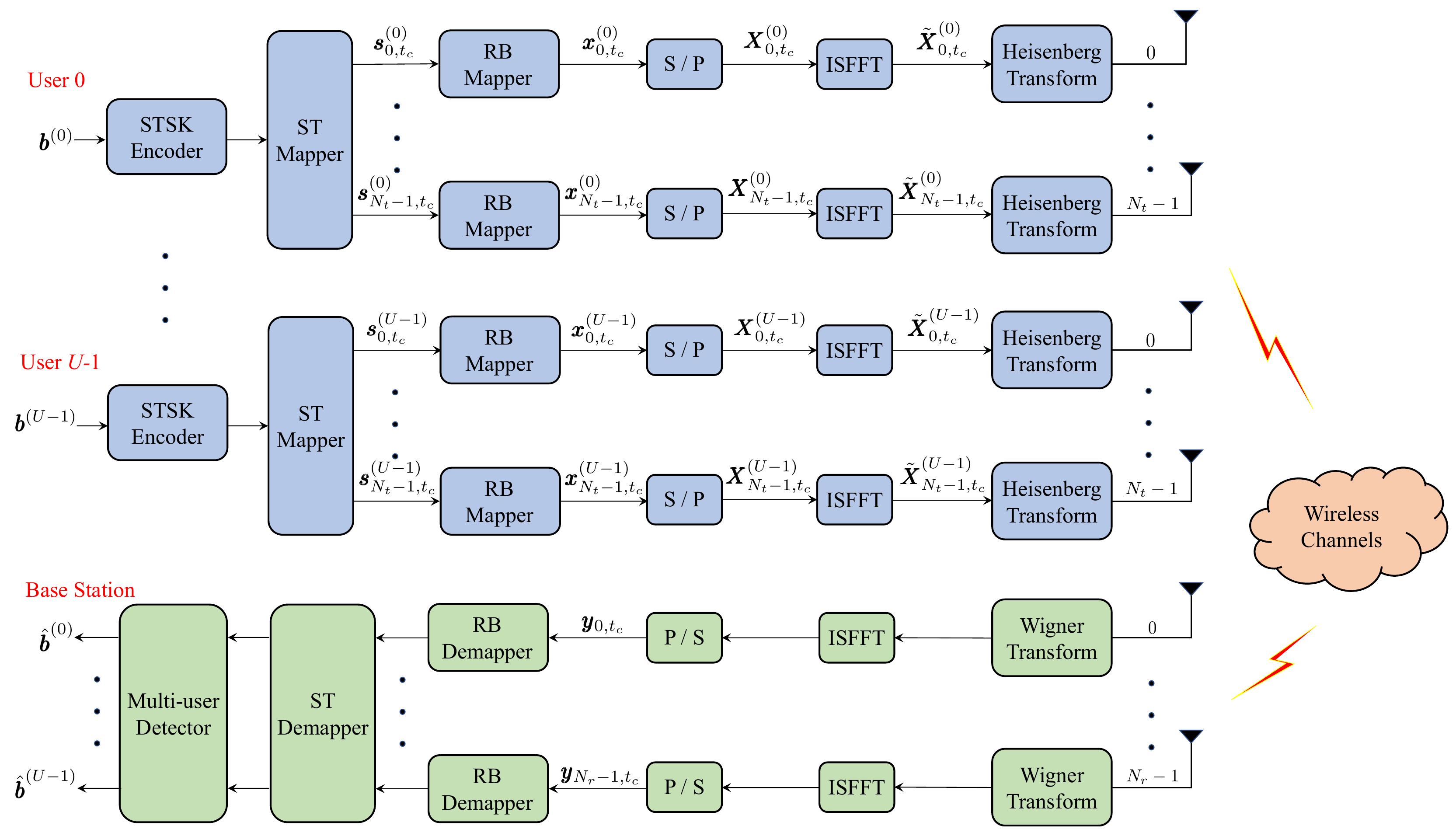}
\caption{Illustration of the STSK-OTFS-MA system.}
\label{Figure1}
\vspace{-1em}
\end{figure*}

Let us consider a single-cell uplink communication scenario, where the information signals of $U$ users are simultaneously transmitted to a base station (BS). Specifically, we assume that $N_t$ TAs are employed by each user, and the BS uses $N_r$ receive antennas (RAs). Moreover, each TA transmits an OTFS signal having the bandwidth of $B=M\Delta f$ and time-slot duration of $T_f=NT$, where $M$ and $N$ denote the number of subcarriers and time intervals within an OTFS time-slot, while $\Delta f$ and $T$ represent the subcarrier spacing and symbol duration, respectively. Hence, we have a total of $M_d=NM$ DD-domain RBs and each user occupies $G=M_d/U$ RBs. As shown in Fig. \ref{Figure1}, the information bit sequence $\pmb{b}^{(u)}\in\mathbb{B}^{\bar{L}}$ transmitted by the user $u$ is first partitioned into $G$ groups, yielding $\pmb{b}^{(u)}=[\pmb{b}^{(u)}_1,\ldots,\pmb{b}^{(u)}_G]$. The $g$th bit sequence $\pmb{b}^{(u)}_g\in\mathbb{B}^{L_b}$, $g=0,\ldots,G-1$, contains $L_b=\bar{L}/G=L_1+L_2$ bits. Explicitly, the subsequence $\pmb{b}^{(u)}_{1,g}\in\mathbb{B}^{L_1}$ is mapped into an index symbol in $\{1,\ldots,Q\}$ for selecting an active DM in $\mathcal{A}=\{\pmb{A}_1,\ldots,\pmb{A}_Q\}$, where we have $L_1=\log_2 Q$. The remaining $L_2$-bit sequences $\pmb{b}^{(u)}_{2,g}\in\mathbb{B}^{L_2}$ are mapped into the normalized quadrature amplitude modulation (QAM)/phase-shift keying (PSK) symbols chosen from the constellation $\mathcal{F}=\{f_1,\ldots,f_V\}$, where $L_2=\log_2 V$. Hence, in an OTFS frame, the total number of bits transmitted per user can be given by $L=UGL_b=NM\log_2 (VQ)$. Assuming that the STSK symbol duration includes $T_c$ OTFS time-slots, the ST codeword {matrices} of {the user $u$} can be expressed as \cite{5599264}
\begin{align}\label{Eq1}	
\pmb{S}_{d,g}^{(u)}=f^{(u)}_{l_g}\pmb{A}^{(u)}_{d,g}\in\mathbb{C}^{N_t\times T_c},
	\end{align}
where $f_{l_g}^{(u)}\subset\mathcal{F}$ and $\pmb{A}^{(u)}_{d,g}\subset\mathcal{A}$ denote a single QAM/PSK symbol and an active DM, respectively. By introducing $\breve{\pmb{s}}_{d,g}^{(u)}=\text{vec}(\pmb{S}_{d,g}^{(u)})$ and stacking all the $G$ RBs of user $u$, the $u$th ST {stacked} codeword vector is formulated as
\begin{align}\label{Eqnew1}
\breve{\pmb{s}}_d^{(u)}= \text{sta}\{\breve{\pmb{s}}_{d,g}^{(u)}\}|_{g=0}^{G-1},\quad g=0,\ldots,G-1.	
\end{align}
Let us parameterize the STSK-OTFS-MA system by the five-tupple $(N_t,N_r,T_c,Q,V)$. Note that the DM set can be generated using diverse design criteria, for example by minimizing the pairwise error probability or by maximizing the DCMC capacity \cite{6316188}, which will be further investigated in Section \ref{section4-4}. In the STSK pre-processing block, the DMs are assumed to be normalized to maintain the transmitted power, and the constraint is given by \cite{5599264}
\begin{align}\label{Eq2}	
	\text{tr}(\pmb{A}_q^H\pmb{A}_q)=T_c,\quad q=1,\ldots,Q.
\end{align}
\begin{figure*}[t]
\centering
\includegraphics[width=0.9\linewidth]{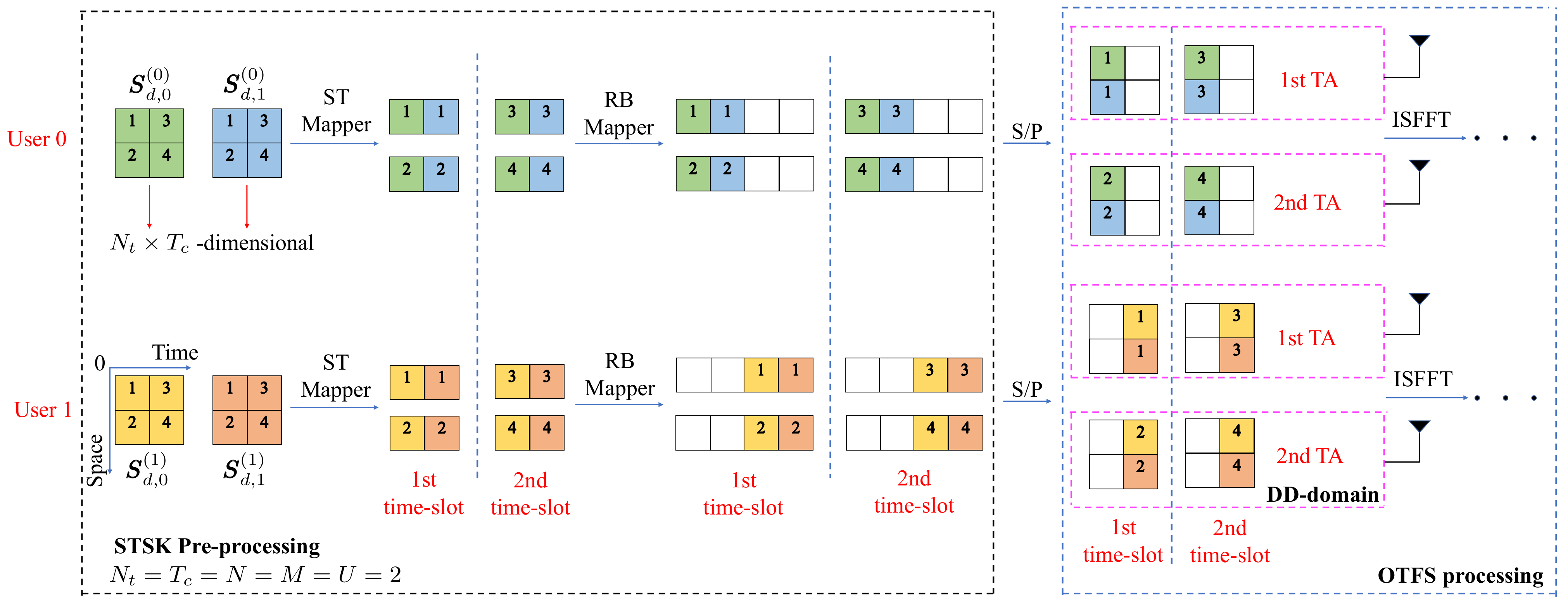}
\caption{A toy example of the STSK-OTFS-MA system with $N_t=T_c=N=M=U=2$.}
\label{Figure1-1}
\end{figure*}
As shown in Fig. \ref{Figure1-1}, the ST codewords are fed into the ST mapper, which is detailed in Section \ref{Section2-2}, and the $u$th user's transmitted frame {output by the ST mapper} $\pmb{s}^{(u)}\in\mathbb{C}^{N_tG\times T_c}$ can be formulated as
\begin{align}\label{Eq5}	
\pmb{s}^{(u)}=\begin{bmatrix}
		\pmb{s}^{(u)}_{0,0} & \cdots & \pmb{s}^{(u)}_{0,T_c-1} \\
		\vdots & \ddots & \vdots \\				
		\pmb{s}^{(u)}_{N_t-1,0} & \cdots & \pmb{s}^{(u)}_{N_t-1,T_c-1}	
		\end{bmatrix},
\end{align}
where we have $\pmb{s}^{(u)}_{n_t,t_c}=[S_{d,0}^{(u)}(n_t,t_c),\ldots,S_{d,G-1}^{(u)}(n_t,t_c)]^T\in\mathbb{C}^{G\times 1}$ for $n_t=0,\ldots,N_t-1$ and $t_c=0,\ldots,T_c-1$, which is formulated based on \eqref{Eq1}. Then the {$G$ ST codeword elements of $\pmb{s}^{(u)}_{n_t,t_c}$} are mapped to $M_d$ RBs, yielding
\begin{align}\label{Eqnew2}
\pmb{x}^{(u)}_{n_t,t_c}=\pmb{\mathcal{P}}^{(u)}\pmb{s}^{(u)}_{n_t,t_c}=[x^{(u)}_{n_t,t_c}(0),\ldots,x^{(u)}_{n_t,t_c}(M_d-1)]^T,	
\end{align}
where $\pmb{\mathcal{P}}^{(u)}$ is the $(M_d\times G)$-element resource allocation matrix. To alleviate the MUI caused by ICI, {we introduce our delay-domain index-based RB allocation scheme, i.e., Scheme 1 illustrated in Fig. \ref{Figure2} \subref{Figure2-1}, where each user occupies $J=M/U$ columns of the DD-domain grids. By contrast, the Doppler-domain index-based resource allocation scheme (Scheme 2) shown in  Fig. \ref{Figure2} \subref{Figure2-2} is invoked as the benchmark.} Let us denote the column indices of user $u$ by $\mathcal{L}^{(u)}=\{l_0^{(u)},\ldots,l_{J-1}^{(u)}\}$. More specifically, if we assume that the ST codewords of user $u$ are assigned to the RBs set $\mathcal{N}^{(u)}$, then the corresponding elements of $\pmb{\mathcal{P}}^{(u)}$ can be expressed as 
\begin{align}\label{Eq7}
	\mathcal{P}^{(u)}(m_d,g)=\begin{cases}
		1, & \mbox{if } m_d\in\mathcal{N}^{(u)}\\
		0, & \mbox{otherwise}
	\end{cases}
\end{align}
for $0\leq m_d\leq M_d-1$. The indices in $\mathcal{N}^{(u)}$ are given by $m_d=l_j^{(u)}N+n$ and $g=j^{(u)}N+n$, where $l_j^{(u)}=(J\times u)+j$ for $n=0,\ldots,N-1$ and $j=0,\ldots,J-1$.
\begin{figure*}[htbp]
\centering
\vspace{-0.6cm}
\subfigure[]{\label{Figure2-1}\includegraphics[width=0.4\linewidth]{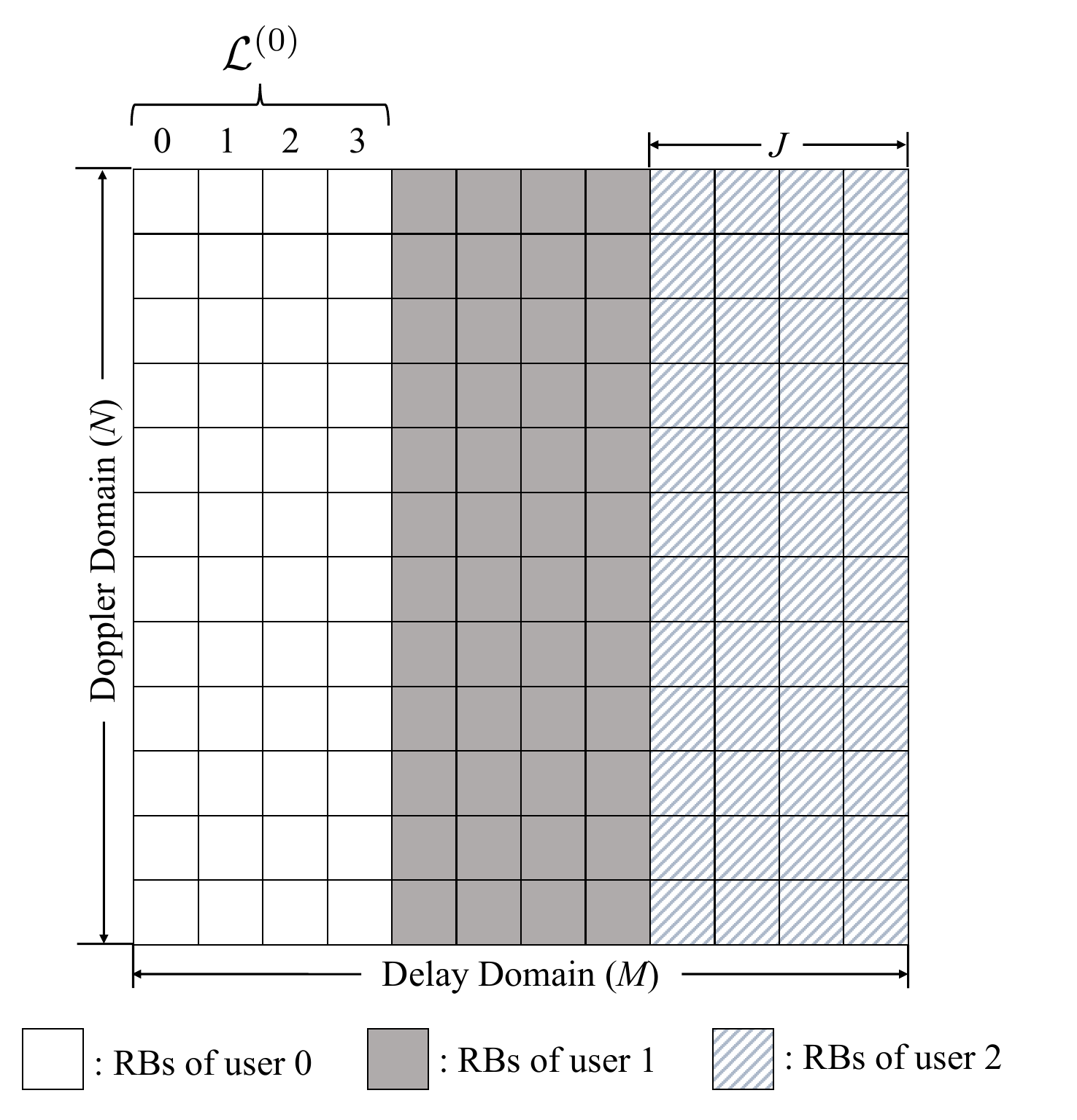}}
\subfigure[]{\label{Figure2-2}\includegraphics[width=0.4\linewidth]{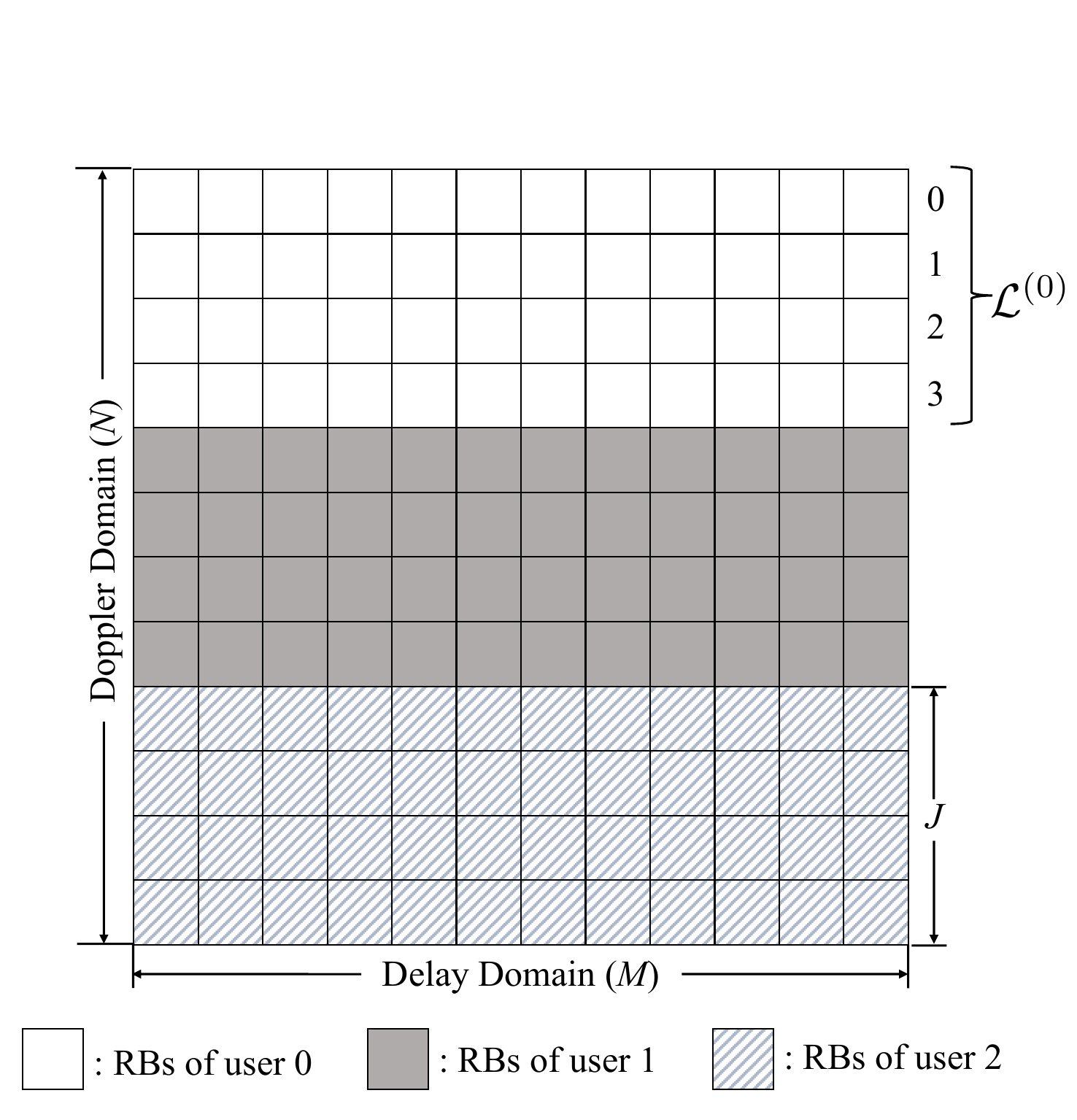}}
\caption{Illustration of resource allocations \subref{Figure2-1} Scheme 1 \subref{Figure2-2} Scheme 2 with $M=12$, $N=12$, $U=3$ and $J=M/U=4$, where $\mathcal{L}^{(u)}$ represents the column index set of user $u$ for $u=0,\ldots,U-1$.}
\vspace{-1.5em}
\label{Figure2}
\end{figure*}

When considering the OTFS processing block, by defining a DD-domain codeword matrix as $\pmb{X}^{(u)}_{n_t,t_c}=\text{vec}^{-1}(\pmb{x}^{(u)}_{n_t,t_c})$, the TF-domain signal can be formulated using ISFFT as
\begin{align}\label{Eq9}
	\tilde{X}^{(u)}_{n_t,t_c}(n,m)=\sum_{k=0}^{N-1}\sum_{l=0}^{M-1}\frac{X^{(u)}_{n_t,t_c}(k,l)}{\sqrt{M_d}}e^{j2\pi\left(\frac{nk}{N}-\frac{ml}{M}\right)},
	\end{align}
for $n=0,\ldots,N-1$ and $m=0,\ldots,M-1$. The transmitted TD signal is obtained by exploiting the Heisenberg transform, yielding
\begin{align}\label{Eq10}
	\tilde{s}^{(u)}_{n_t,t_c}(t)=\sum_{n=0}^{N-1}\sum_{m=0}^{M-1}\tilde{X}^{(u)}_{n_t,t_c}(n,m)g_{\text{tx}}(t-nT)e^{j2\pi m\Delta f(t-nT)},
	\end{align}
where $g_{\text{tx}}(t)$ denotes the transmit waveform.
\subsection{Received Signals}\label{Section2-2}
Assuming that the synchronization among uplink users is perfect. Let us consider a $P$-path DD-domain time-varying multipath channel model between the $n_t$th TA of user $u$ and the $n_r$th RA, which can be formulated as $h^{(u)}_{n_r,n_t}(\tau,\nu)=\sum_{i=1}^{P}h^{(u)}_{i,n_r,n_t}\delta(\tau-\tau_i)\delta(\nu-\nu_i)$, where $h^{(u)}_{i,n_r,n_t}$, $\tau_i$ and $\nu_i$ denote the complex-valued path gain between the $n_t$th TA and $n_r$th RA, normalized delay and Doppler shifts introduced by the $i$th path, respectively \cite{7925924}. Here we have $h_i\sim\mathcal{CN}(0,1/P),\forall i$, which is independent of $\tau_i$ and $\nu_i$. Therefore, the delay and Doppler shifts corresponding to the $i$th reflector are given by $\nu_i=\frac{k_i}{NT},\tau_i=\frac{l_i}{M\Delta f}$, {where $l_i=a_i+\alpha_i$ and $k_i=b_i+\beta_i$ represent the normalized delay and Doppler indices associated with the $i$th path, where $a_i$ and $b_i$ denote the integer delay and Doppler indices, while the fractional components are given by $\alpha_i,\beta_i\in\mathcal{U}[-\frac{1}{2},\frac{1}{2}]$.} During the $t_c$th OTFS time-slot, the TD signal of the $n_r$th RA received from the $n_t$th TA of user $u$ can be expressed as \cite{8424569}
\begin{align}
r^{(u)}_{n_r,n_t,t_c}(t)=&\int\int h^{(u)}_{n_r,n_t}(\tau,\nu)\tilde{s}^{(u)}_{n_t,t_c}(t-\tau)e^{j2\pi \nu(t-\tau)}d\tau d\nu+n^{(u)}_{n_r,n_t,t_c}(t),	
\end{align}
where $n^{(u)}_{n_r,n_t,t_c}(t)$ is the complex-valued additive white Gaussian noise (AWGN). Based on the Wigner transform, the elements of the corresponding received TF-domain codeword matrix $\tilde{\pmb{Y}}^{(u)}_{n_r,n_t,t_c}\in\mathbb{C}^{N\times M}$ can be obtained as
\begin{align}\label{Eq14}
	\tilde{Y}^{(u)}_{n_r,n_t,t_c}(n,m)=\int r^{(u)}_{n_r,n_t}(t')g_{\text{rx}}^{*}(t'-nT)e^{-j2\pi m\Delta f(t'-nT)}dt',	
	\end{align}
where $g_{\text{rx}}(t)$ is the receive waveform. Then, upon utilizing the SFFT, the received DD-domain codeword matrix can be formulated as
\begin{align}\label{Eq15}
	Y^{(u)}_{n_r,n_t,t_c}(k,l)=\sum_{n=0}^{N-1}\sum_{m=0}^{M-1}\frac{\tilde{Y}^{(u)}_{n_r,n_t,t_c}(n,m)}{\sqrt{M_d}}e^{-j2\pi\left(\frac{nk}{N}-\frac{ml}{M}\right)},
	\end{align}
for $k=0,\ldots,N-1$ and $l=0,\ldots,M-1$. Assuming that both the transmit and receive waveforms satisfy the bi-orthogonal condition, the vector-form input-output relationship for user $u$ can be formulated as \cite{8424569}
\begin{align}\label{Eq17}
	\pmb{y}^{(u)}_{n_r,n_t,t_c}=\pmb{H}^{(u)}_{n_r,n_t}\pmb{x}^{(u)}_{n_t,t_c}+\pmb{n}^{(u)}_{n_r,n_t,t_c},
	\end{align}
where we have $\pmb{y}^{(u)}_{n_r,n_t,t_c}=\text{vec}\left(\pmb{Y}^{(u)}_{n_r,n_t,t_c}\right)$, and $\pmb{n}^{(u)}_{n_r,n_t,t_c}$ denotes the complex-valued AWGN vector. Moreover, the effective DD-domain channel matrix $\pmb{H}^{(u)}_{n_r,n_t}$ can be expressed as $\pmb{H}^{(u)}_{n_r,n_t}=\sum_{i=1}^{P}\pmb{I}_M(l_i)\otimes\left[\pmb{I}_N(k_i)h^{(u)}_{i,n_r,n_t}e^{-j2\pi\frac{l_ik_i}{M_d}}\right]$ \cite{liu2021message}. Let $\pmb{H}^{(u)}_{n_r}=\left[\pmb{H}^{(u)}_{0,n_r},\ldots,\pmb{H}^{(u)}_{N_t-1,n_r}\right]$ and $\pmb{x}^{(u)}_{t_c}=\text{sta}\{\pmb{x}^{(u)}_{n_t,t_c}\}|_{n_t=0}^{N_t-1}$. Then the DD-domain codeword vector received by the $n_r$th RA {within the $t_c$th time-slot} can be expressed as
\begin{align}\label{Eq19}
	\pmb{y}_{n_r,t_c}&=\sum_{u=0}^{U-1}\pmb{y}^{(u)}_{n_r,t_c}+\pmb{n}_{n_r,t_c}=\sum_{u=0}^{U-1}\sum_{n_t=0}^{N_t-1}\pmb{H}^{(u)}_{n_r,n_t}\pmb{x}^{(u)}_{n_t,t_c}+\pmb{n}_{n_r,t_c}\nonumber\\
	&=\sum_{u=0}^{U-1}\pmb{H}^{(u)}_{n_r}\pmb{x}^{(u)}_{t_c}+\pmb{n}_{n_r,t_c},
	\end{align}
where $\pmb{n}_{n_r,t_c}$ is the corresponding complex AWGN vector with a zero mean and a covariance matrix of $N_0\pmb{I}_{M_d}$, expressed as $\mathcal{CN}(0,N_0\pmb{I}_{M_d})$. Hence, the average SNR per RA is given by $\gamma=1/N_0$. {By invoking all the signals of $N_r$ RAs and} let $\bar{\pmb{y}}_{t_c}=\text{sta}\{\pmb{y}_{n_r,t_c}\}_{n_r=0}^{N_r-1}$, then it can be shown that the end-to-end DD-domain input-output relationship for the time-slot $t_c$ is given as
\begin{align}\label{Eq21}
	\bar{\pmb{y}}_{t_c}=\sum_{u=0}^{U-1}\bar{\pmb{H}}^{(u)}\pmb{x}^{(u)}_{t_c}+\tilde{\pmb{n}}_{t_c},
\end{align}
for $t_c=0,\ldots,T_c-1$, where $\tilde{\pmb{n}}_{t_c}=\text{sta}\{\pmb{n}_{n_r,t_c}\}|_{n_r=0}^{N_r-1}$ denotes the stacked noise vector at the BS side. The DD-domain MIMO channel matrix of the $u$th user can be expressed as $\bar{\pmb{H}}^{(u)}=\text{sta}\{\pmb{H}_{n_r}^{(u)}\}|_{n_r=0}^{N_r-1}$. Moreover, we invoke the relationship between the ST codewords and the stacked codeword vector {shown in \eqref{Eqnew2},} the transmitted codeword vector $\pmb{x}^{(u)}_{t_c}$ can be formulated as
\begin{align}\label{Eq22}
	\pmb{x}^{(u)}_{t_c}&=\left[\left(\pmb{\mathcal{P}}^{(u)}\pmb{s}^{(u)}_{0,t_c}\right)^T,\ldots,\left(\pmb{\mathcal{P}}^{(u)}\pmb{s}^{(u)}_{N_t-1,t_c}\right)^T\right]^T\nonumber\\
	&=\left(\pmb{I}_{N_t}\otimes\pmb{\mathcal{P}}^{(u)}\right)\left[\left(\pmb{s}^{(u)}_{0,t_c}\right)^T,\ldots,\left(\pmb{s}^{(u)}_{N_t-1,t_c}\right)^T\right]^T\nonumber\\
	&=\bar{\pmb{\mathcal{P}}}^{(u)}\pmb{s}^{(u)}_{t_c},
	\end{align}
where $\bar{\pmb{\mathcal{P}}}^{(u)}$ is the equivalent resource allocation matrix of user $u$. Then, upon applying \eqref{Eq22} to \eqref{Eq21}, $\bar{\pmb{y}}_{t_c}$ can be rewritten as
\begin{align}\label{Eq23}
	\bar{\pmb{y}}_{t_c}&=\sum_{u=0}^{U-1}\bar{\pmb{H}}^{(u)}\bar{\pmb{\mathcal{P}}}^{(u)}\pmb{s}^{(u)}_{t_c}+\tilde{\pmb{n}}_{t_c}\nonumber\\
	&=\sum_{u=0}^{U-1}\pmb{\Omega}^{(u)}\pmb{s}^{(u)}_{t_c}+\tilde{\pmb{n}}_{t_c},\quad t_c=0,\ldots,T_c-1,
	\end{align}
where $\pmb{\Omega}^{(u)}=\bar{\pmb{H}}^{(u)}\bar{\pmb{\mathcal{P}}}^{(u)}$ is a $(M_dN_r\times GN_t)$-dimensional matrix.

{By considering all the $T_c$ time-slots,} let us define the overall received symbol vector and the transmitted STSK symbol vector of user $u$ as $\tilde{\pmb{y}}=\text{sta}\{\bar{\pmb{y}}_{t_c}\}|_{t_c=0}^{T_c-1}$ and $\bar{\pmb{s}}^{(u)}=\text{sta}\{\pmb{s}^{(u)}_{t_c}\}|_{t_c=0}^{T_c-1}$, respectively. Consequently, $\tilde{\pmb{y}}$ can be obtained as
\begin{align}\label{Eq24}
	\tilde{\pmb{y}}&=\sum_{u=0}^{U-1}\left(\pmb{I}_{T_c}\otimes\pmb{\Omega}^{(u)}\right)\bar{\pmb{s}}^{(u)}+\tilde{\pmb{n}}\nonumber\\
	&=\sum_{u=0}^{U-1}\tilde{\pmb{\Omega}}^{(u)}\bar{\pmb{s}}^{(u)}+\tilde{\pmb{n}}=\tilde{\pmb{\Omega}}\tilde{\pmb{s}}+\tilde{\pmb{n}},
	\end{align}
where $\tilde{\pmb{\Omega}}=\left[\tilde{\pmb{\Omega}}^{(0)},\ldots,\tilde{\pmb{\Omega}}^{(U-1)}\right]\in\mathbb{C}^{M_dN_rT_c\times GN_tT_cU}$, $\tilde{\pmb{s}}=\text{sta}\{\bar{\pmb{s}}^{(u)}\}|_{u=0}^{U-1}\in\mathbb{C}^{GN_tT_cU\times 1}$ and $\tilde{\pmb{n}}=\text{sta}\{\tilde{\pmb{n}}_{t_c}\}|_{t_c=0}^{T_c-1}$. 

{Now we further detail the ST mapper shown in Fig. \ref{Figure1}. For deriving the associated input-output relationship, we stack the transmitted ST codewords of $U$ users shown in \eqref{Eqnew1}, yielding $\tilde{\pmb{s}}_d=\text{sta}\{\breve{\pmb{s}}_d^{(u)}\}|_{u=0}^{U-1}$. As illustrated in Fig. \ref{Figure1} and Fig. \ref{Figure1-1} of Section \ref{Section2-1}, the symbol vector $\tilde{\pmb{s}}_d$ is essentially obtained by rearranging all the $GN_tT_cU$ ST codeword elements $S^{(u)}_{d,g}(n_t,t_c)$ based on the following \underline{order 1} of \emph{1) TA index 2) time-slot index 3) RB index 4) user index}. However, according to the STSK and multiuser system design principles \cite{5599264,10129061,7448828,8515088} and to the transceiver structure of Section \ref{Section2-1} and Section \ref{Section2-2}, the above-mentioned ST codeword elements should be placed in the following \underline{order 2} of \emph{1) RB index 2) TA index 3) user index 4) time-slot index}, which can be further verified based on our derivation of \eqref{Eq17}-\eqref{Eq24}.} Therefore, we have $\tilde{\pmb{s}}=\pmb{\Upsilon}\tilde{\pmb{s}}_d$, where $\pmb{\Upsilon}$ is the $(GN_tT_cU\times GN_tT_cU)$-element ST mapping matrix, whose elements are defined as
\begin{align}\label{Eq26}
	\Upsilon(d_x,d_y)=\begin{cases}
		1, & \mbox{if } d_x=g+n_tG+uN_tG+t_cGUN_t\\
		   & \mbox{and } d_y=n_t+t_cN_t+gN_tT_c+uGN_tT_c\\
		0, & \mbox{otherwise},  
	\end{cases}
\end{align}
for $0\leq d_x,d_y\leq GN_tUT_c-1$. {It should be noted that $d_x$ and $d_y$ are formulated based on \underline{order 2} and \underline{order 1}, respectively.} Finally, based on \eqref{Eq24} and \eqref{Eq26},  the end-to-end input-output relationship for the OTFS frame can be expressed as
\begin{align}\label{Eq27}
	\tilde{\pmb{y}}&=\tilde{\pmb{\Omega}}\pmb{\Upsilon}\tilde{\pmb{s}}_d+\tilde{\pmb{n}}=\tilde{\pmb{\Omega}}\pmb{\Upsilon}\bar{\pmb{\chi}}\pmb{K}+\tilde{\pmb{n}}=\pmb{C}\pmb{K}+\tilde{\pmb{n}},
\end{align}
where $\bar{\pmb{\chi}}=\pmb{I}_{UG}\otimes\pmb{\chi}$ with $\pmb{\chi}=[\text{vec}(\pmb{A}_1),\ldots,\text{vec}(\pmb{A}_Q)]\in\mathbb{C}^{N_tT_c\times Q}$. Moreover, the equivalent transmitted symbol vector can be defined as $\pmb{K}=\text{sta}\{\pmb{K}^{(u)}\}_{u=0}^{U-1}$, in which the sub-vectors can be formulated as $\pmb{K}^{(u)}=\text{sta}\{\bar{\pmb{K}}_{\mathcal{I}_{g}^{(u)}}\}|_{g=0}^{G-1}$, and the index sets are given by $\mathcal{I}_g^{(u)}=\{q_g^{(u)},l_g^{(u)}\}$, where $q_g^{(u)}\in\{1,\ldots,Q\}$ and $l_g^{(u)}\in\{1,\ldots,V\}$ for $u=0,\ldots,U-1$ and $g=0,\ldots,G-1$. Moreover, the $g$th equivalent transmitted symbol vector of the $u$th user can be formulated as
\begin{align}\label{Eq29}
	\bar{\pmb{K}}_{\mathcal{I}_{g}^{(u)}}=[\underbrace{0, \cdots, 0}_{q_g^{(u)}-1}, f_{l_g^{(u)}}, \underbrace{0, \cdots, 0}_{Q-q_g^{(u)}}]^T\in\mathbb{C}^{Q\times 1},
\end{align}
where the $l_g^{(u)}$th QAM/PSK symbol $f_{l_g^{(u)}}$ is located in the $q_g^{(u)}$th element, while the active DM in the $g$th STSK block of user $u$ is denoted as $\pmb{A}_{q_g^{(u)}}$. According to \eqref{Eq27}-\eqref{Eq29}, it can be observed that a total of $M_d$ ST codewords are transmitted in the entire system, hence only $M_d$ elements of $\tilde{\pmb{K}}$ have a non-zero value. Therefore, the index candidate sets of non-zero valued elements are represented as $\mathcal{Q}=\left\{\mathcal{Q}_1,\ldots,\mathcal{Q}_C\right\}$, which has $C=2^{M_dL_1}=Q^{M_d}$ index candidate subsets; The $c$th subset can be expressed as $\mathcal{Q}_c=\left\{\mathcal{Q}_c(0),\ldots,\mathcal{Q}_c(M_d-1)\right\}\subset\mathcal{Q}$, whose elements obey $\mathcal{Q}_c(m_d)\in\mathbb{Z}^{QM_d}_{+}$ for $m_d=0,\ldots,M_d-1$ and $c = 1,\ldots,C$. For a given $\pmb{K}$, the index candidate subset is denoted as $\mathcal{I}$, where we have $\mathcal{I}=\mathcal{Q}_c\subset\mathcal{Q}$, and the APM symbols can be expressed as $\pmb{K}_d=[K_d(0),\ldots,K_d(M_d-1)]^T\in\mathcal{F}^{M_d\times 1}$. For the sake of demonstration, all the candidates of the ST codeword vectors can be expressed as a specifically designed codebook, yielding
\begin{align}\label{Eq30}
	\mathcal{B}=\left\{\pmb{B}_1,\ldots,\pmb{B}_{2^L}: \pmb{B}_i\in\mathbb{C}^{QM_d},i=1,\ldots,2^L\right\},
\end{align}
which will be discussed in Section \ref{Section4}, while $\pmb{K}$ is a vector having elements selected from $\mathcal{B}$. Based on \eqref{Eq27}, the received symbol vector $\tilde{\pmb{y}}$ for a given $\pmb{K}$ follows the Gaussian probability density function (PDF) of
\begin{align}\label{Eq31}
	p(\tilde{\pmb{y}}|\pmb{K})=\frac{1}{(\pi N_0)^{M_dN_rT_c}}\exp\left(-\frac{||\tilde{\pmb{y}}-\pmb{C}\pmb{K}||^2}{N_0}\right).
\end{align}
{Finally, the attainable rate can be formulated as} 
\begin{align}
	R=\frac{L}{NMT_c}=\frac{\log_2 (VQ)}{T_c}\ \text{bits/s/Hz}.
\end{align}
\vspace{-2em}
\section{Multiuser Detection in STSK-OTFS-MA Systems}\label{Section3}
In this section, we first introduce the optimal maximum \emph{a posteriori} detector (MAPD) or MLD designed for our STSK-OTFS-MA system. Typically, the complexity of the optimal MLD becomes excessive even for moderate constellations. To mitigate this problem, a pair of detectors are proposed. Specifically, we commence by deriving a low-complexity PRCGD, and then we detail the IRCD. Finally, the complexity analysis of the proposed detectors is provided.
\subsection{Maximum A Posteriori Detector (MAPD)}\label{section3-1}
Given the conditional PDF in \eqref{Eq31}, the optimum MAPD maximizes the \emph{a posteriori} probability of the equivalent transmitted vector $\pmb{K}$, yielding $\pmb{K}^{\text{MAP}}=\underset{\pmb{B}_i\in\mathcal{B}}\argmax\left\{p(\pmb{B}_i|\tilde{\pmb{y}})\right\}$. Assuming that the mapping process of different candidates in $\mathcal{B}$ is independent and equiprobable, the MAPD is equivalent to the MLD, which can be written as
\begin{align}\label{Eq34}
	\pmb{K}^{\text{ML}}=\underset{\pmb{B}_i\in\mathcal{B}}{\arg\min} \left\{||\tilde{\pmb{y}}-\pmb{C}\pmb{B}_i||^2\right\}.
	\end{align}
\subsection{Progressive Residual Check Greedy Detector (PRCGD)}\label{section3-2}
We firstly rewrite the input-output relationship of \eqref{Eq27} as
\begin{align}\label{Eq35}
	\tilde{\pmb{y}}&=\pmb{C}\pmb{K}+\tilde{\pmb{n}}=\pmb{C}\pmb{\Upsilon}_{\mathcal{I}}\pmb{K}_d+\tilde{\pmb{n}}=\pmb{C}_{\mathcal{I}}\pmb{K}_d+\tilde{\pmb{n}},	
	\end{align}
where $\pmb{C}_{\mathcal{I}}=\pmb{C}\pmb{\Upsilon}_{\mathcal{I}}\in\mathbb{C}^{M_dN_rT_c\times M_d}$ and $\pmb{\Upsilon}_{\mathcal{I}}$ is a $(QM_d\times M_d)$-element mapping matrix associated with ${\mathcal{I}}$. Assuming that the candidates in the index candidate sets $\mathcal{Q}$ and in the constellation set $\mathcal{F}$ are independent and equiprobable, similar to \eqref{Eq34}, the corresponding joint MLD can be formulated as $\left(\mathcal{I}^{\text{ML}},\pmb{K}_d^{\text{ML}}\right)=\underset{\mathcal{Q}_c\subset\mathcal{Q},\pmb{f}\in\mathcal{F}^{M_d}}\argmin\left\{\left\|\tilde{\pmb{y}}-\pmb{C}_{\mathcal{Q}_c}\pmb{f}\right\|^2\right\}$. Since the complexity of the MLD is excessive, we propose the PRCGD. Based on the sparse structure of the transmitted symbol vector $\pmb{K}$, our objective is to harness the philosophy of greedy algorithms \cite{8478829}, which are often employed for sparse recovery in low-complexity multiuser detection (MUD). Specifically, the proposed PRCGD can provide locally optimal detection results based on the elements of $\pmb{K}$. Moreover, our PRCGD includes the reliability sorting and progressive detection stages detailed below.
\begin{algorithm}[htbp]
\footnotesize
\caption{Progressive Residual Check Greedy Detector}
\label{alg1}
\begin{algorithmic}[1]
    \Require
      $\tilde{\pmb{y}}$, $\pmb{C}$, $\mathcal{Q}$, $\gamma_s$ and $\epsilon_0$.
      \State \textbf{Preparation}: Set the maximum number of iteration $T_1$, $\epsilon_{\infty}=\infty$, $\mathcal{I}^{\text{PRCGD}}=\emptyset$ and $\pmb{K}_d^{\text{PRCGD}}=\emptyset$.
    \State $// \text{Reliability Sorting:}$    
    \State Employ LMMSE detection according as
    \State $\tilde{\pmb{K}}=\left(\pmb{C}^H\pmb{C}+\frac{1}{\gamma_s}\pmb{I}_{QM_d}\right)^{-1}\pmb{C}^H\tilde{\pmb{y}}$.
    \State Obtain the measurements of the index reliabilities as
    \State $\mathcal{J}=\{j_1,\ldots,j_{QM_d}\}\quad\text{subject to }\left|\tilde{K}(j_1)\right|^2\geqslant\ldots\geqslant\left|\tilde{K}(j_{QM_d})\right|^2$.
    \State $// \text{Progressive Detection:}$  
    \For{$t=1$ to $T_1$}
    \State Collect the DAPs $\mathcal{Q}^t=\left\{\mathcal{Q}_1^t,\ldots,\mathcal{Q}_{C_t}^t\right\}$, 
    \Statex \quad\ where $\bigcap_{c_t=1}^{C_t}\mathcal{Q}_{c_t}^t=j_t$.
    \If{$\epsilon^t<\epsilon_0$}
    \BREAK
    \Else
    \For{$c_t=1$ to $C_t$}
    \State Compute the least square solution as:
    \State $\hat{\pmb{K}}_{c_t,d}=\pmb{C}_{\mathcal{Q}_{c_t}^t}^{\dagger}\tilde{\pmb{y}}$
    \State Carry out APM symbol estimation as
    \State $f^t_{c_t}(j)=\underset{f_v\in\mathcal{F}}\argmin\left|\hat{K}_{c_t,d}(j)-f_v\right|^2$,
    \Statex \quad\quad\quad for $m_d=0,\ldots,M_d-1$.
    \EndFor
    \State Obtain the local optimal set as
    \State $\left(\mathcal{I}^t,\pmb{K}_d^t\right)=\underset{\mathcal{Q}_{c_t}^{t}\subset\mathcal{Q}^t,\pmb{f}_{c_t}^t\in\hat{\mathcal{F}}^t}\argmin\left\{\left\|\tilde{\pmb{y}}-\pmb{C}_{\mathcal{Q}_{c_t}^{t}}\pmb{f}_{c_t}^{t}\right\|^2\right\}$.
    \State Compute the residual error as
	\State $\epsilon^t=\left\|\tilde{\pmb{y}}-\pmb{C}_{\mathcal{I}^t}\pmb{K}_d^t\right\|^2$.
	\If{$\epsilon^t<\epsilon_0$}
	\State $\mathcal{I}^{\text{PRCGD}}=\mathcal{I}^t$ and $\pmb{K}_d^{\text{PRCGD}}=\pmb{K}_d^t$
	\BREAK
	\ElsIf{$\epsilon^t<\epsilon_{\infty}$}
	\State $\mathcal{Q}\leftarrow\mathcal{Q}\setminus\mathcal{Q}^t$, $\epsilon_{\infty}=\epsilon^{t}$, $\mathcal{I}^{\text{PRCGD}}=\mathcal{I}^t$ and
	\Statex \quad\quad\quad $\pmb{K}_d^{\text{PRCGD}}=\pmb{K}_d^t$
	\Else	
	\EndIf
	\EndIf
    \EndFor
\State \textbf{Output} $\mathcal{I}^{\text{PRCGD}}=\mathcal{I}^t$ and $\pmb{K}_d^{\text{PRCGD}}=\pmb{K}_d^t$.
\end{algorithmic}
\end{algorithm}

At the reliability sorting stage, the received symbol vector $\tilde{\pmb{y}}$ is first processed by linear minimum mean square error (LMMSE) estimation to obtain the soft estimates of $\pmb{K}$ as $\tilde{\pmb{K}}=\left(\pmb{C}^H\pmb{C}+\frac{1}{\gamma_s}\pmb{I}_{QM_d}\right)^{-1}\pmb{C}^H\tilde{\pmb{y}}$, where $\tilde{\pmb{K}}=[\tilde{K}(0),\ldots,\tilde{K}(QM_d-1)]^T\in\mathbb{C}^{QM_d\times 1}$ and $\gamma_s=\gamma/Q$ is the generalized average SNR per symbol. The elements in $\tilde{\pmb{K}}$ having relatively high magnitudes should also have relatively high probabilities of being active in $\pmb{K}$, which becomes more pronounced in the high-SNR region. Hence, inspired by \cite{8478829,9507331}, we can order the magnitudes of the elements in $\tilde{\pmb{K}}$ in descending order to reflect the reliability of the index symbols, yielding
\begin{align}\label{Eq39}
	\mathcal{J}=&\{j_1,\ldots,j_{QM_d}\}\quad\text{subject to }\left|\tilde{K}(j_1)\right|^2\geqslant\ldots\geqslant\left|\tilde{K}(j_{QM_d})\right|^2,
\end{align}
where we have $j_l\in\{1,\ldots,QM_d\}$ for $l=1,\ldots,QM_d$ and $j_l\neq j_q,\forall l\neq q$. Then, based on the reliability set $\mathcal{J}$, the PRCGD enters the progressive detection stage, which is detailed below.

During the progressive detection stage, the PRCGD carries out the index symbol detection and the symbol-wise APM symbol detection separately. To elaborate further, we first select $j_t$ from the reliability reliability set $\mathcal{J}$ and exploit $C_t$ DAPs in the $t$th iteration, yielding $\mathcal{Q}^t=\left\{\mathcal{Q}_1^t,\ldots,\mathcal{Q}_{C_t}^t\right\}\subset\mathcal{Q}$, where $\mathcal{Q}_{c_t}^t(m_d)\in\mathbb{Z}^{QM_d}_{+}$ for $m_d=0,\ldots,M_d-1$ and $c_t=1,\ldots,C_t$. Based on the reliability set $\mathcal{J}$, the DAPs are chosen under the constraint that $j_t$ is a common value in all selected subsets, i.e., we have $\bigcap_{c_t=1}^{C_t}\mathcal{Q}_{c_t}^t=j_t$. Upon invoking the DAPs $\mathcal{Q}_{c_t}^t$ as the \emph{a priori} information, the ensuing APM symbol estimation can be formulated as the following optimization problem of $\hat{\pmb{K}}_{c_t,d}=\underset{\pmb{a}\in\mathbb{C}^{M_d\times 1}}\argmin||\tilde{\pmb{y}}-\pmb{C}_{\mathcal{Q}_{c_t}^t\pmb{a}}||^2$. Then the corresponding least square solution can be formulated as
\begin{align}\label{Eq42}
	\hat{\pmb{K}}_{c_t,d}&=\pmb{C}_{\mathcal{Q}_{c_t}^t}^{\dagger}\tilde{\pmb{y}}=\pmb{C}_{\mathcal{Q}_{c_t}^t}^{\dagger}\pmb{C}_{\mathcal{I}}\pmb{K}_d+\pmb{C}_{\mathcal{Q}_{c_t}^t}^{\dagger}\tilde{\pmb{n}}\nonumber\\
	&=\pmb{K}_d+\pmb{r}_{\mathcal{Q}_{c_t}^t,\mathcal{I}}+\bar{\pmb{n}},
\end{align}
where we have the residual interference $\pmb{r}_{\mathcal{Q}_{c_t}^t,\mathcal{I}}=\pmb{0}$ in the case that all the index symbols are detected correctly, i.e., $\mathcal{Q}_{c_t}^t=\mathcal{I}$, and $\bar{\pmb{n}}=\pmb{C}_{\mathcal{Q}_{c_t}^t}^{\dagger}\tilde{\pmb{n}}$ is the corresponding AWGN vector. Based on the DAPs $\mathcal{Q}_{c_t}^t$, the estimates of the APM symbols $\pmb{f}^t_{c_t}=[f^t_{c_t}(0),\ldots,f^t_{c_t}(M_d-1)]^T$ can be obtained by symbol-wise ML detection, yielding,
\begin{align}\label{Eq43}
	f^t_{c_t}(m_d)=\underset{f_v\in\mathcal{F}}\argmin\left|\hat{K}_{c_t,d}(j)-f_v\right|^2,
\end{align}
for $m_d=0,\ldots,M_d-1$ and $v=1,\ldots,V$. Hence, after testing all the $C_t$ DAPs, the PRCGD delivers the corresponding APM candidate sets $\hat{\mathcal{F}}^t=\{\pmb{f}^t_{1},\ldots,\pmb{f}^t_{C_t}\}$. Now we have obtained the estimates of the APM symbols and the DAPs grouped as $\{\pmb{f}_{c_t}^t,\mathcal{Q}_{c_t}^t\}_{c_t=1}^{C_t}$. Therefore, the locally optimal set can be formulated as $\left(\mathcal{I}^t,\pmb{K}_d^t\right)=\underset{\mathcal{Q}_{c_t}^{t}\subset\mathcal{Q}^t,\pmb{f}_{c_t}^t\in\hat{\mathcal{F}}^t}\argmin\left\{\left\|\tilde{\pmb{y}}-\pmb{C}_{\mathcal{Q}_{c_t}^{t}}\pmb{f}_{c_t}^{t}\right\|^2\right\}$, where the residual error can be expressed as $\epsilon^t=\left\|\tilde{\pmb{y}}-\pmb{C}_{\mathcal{I}^t}\pmb{K}_d^t\right\|^2$. Typically, the progressive detection terminates in the case of $\epsilon^t<\epsilon_0$, where $\epsilon_0$ is the predefined termination parameter. However, if this condition cannot be satisfied after testing all DAPs, the PRCGD returns the corresponding set with minimum residual error. The proposed PRCGD is summarized in Algorithm \ref{alg1}.
\vspace{-1em}
\subsection{Iterative Reduced-Space Check Detector (IRCD)}\label{section3-3}
The main philosophy of IRCD is to carry out the detection of DAPs and APM symbols separately. Then near-ML detection performance can be obtained by only testing a reduced set of the entire DAP space $\mathcal{Q}$. In contrast to the PRCGD, the index reliabilities of all elements in $\tilde{\pmb{K}}$ are invoked in the IRCD. Explicitly, the IRCD first derives the index reliability metric based on each DAP $\mathcal{Q}_c$ after employing the LMMSE detection, yielding $\rho_c = \sum_{m_d=0}^{M_d-1}\left|\tilde{K}(i^c_{m_d})\right|^2$ for $i^c_{m_d}=\mathcal{Q}_c(m_d)\in\{0,\ldots,QM_d-1\}$ and $c=1,\ldots,C$. Then the reliability metrics of all $C=Q^{M_d}$ DAPs are sorted in descending order, which can be formulated as
\begin{align}\label{Eq45-1}
	\mathcal{R}=\{{i_1},\ldots,{i_C}\}\quad\text{subject to }{\rho}_{i_1}\geqslant\ldots\geqslant{\rho}_{i_C},
\end{align}
where we have $i_c\in\{1,\ldots,C\}$ for $c=1,\ldots,C$ and ${\rho}_{i_p}\neq{\rho}_{i_q}$, $\forall p\neq q$. Similar to the PRCGD, the DAP associated with a higher value ${\rho}_{i_c}$ in $\mathcal{R}$ can be regarded as the correct result with a higher probability, especially in the high-SNR scenarios. It should be noted that our IRCD first tests the DAPs corresponding to ${\rho}_{i_1}$ with the highest priority in the following stage, where we carry out the detection of the APM symbols, as discussed below.

In the second stage, the proposed IRCD first selects the DAP $\mathcal{I}^t$ based on the reliability metric $\rho_{i_t}$ in the $t$th iteration. Consequently, the detected APM symbols $\pmb{K}_d^t\in\mathcal{F}^{M_d}$ can be obtained based on the least square approach and the symbol-wise ML detection of \eqref{Eq42} and \eqref{Eq43}, respectively. Hence, the detected set and APM symbols in the $t$th iteration can be grouped as ${\mathcal{G}}^t=\{\mathcal{I}^t,\pmb{K}_d^t\}$. Assume that there are $T_2$ DAPs to be tested during the second stage. Hence, the final detected DAP and the APM symbols can be obtained as $(\mathcal{I}^{\text{IRCD}},\pmb{K}_d^{\text{IRCD}})=\underset{(\mathcal{I}^t,\pmb{K}_d^t)\subset\mathcal{G}}\argmin\left\|\tilde{\pmb{y}}-\pmb{C}_{\mathcal{I}^t}\pmb{K}_d^t\right\|^2$, where $\mathcal{G}=\{\mathcal{G}^1\cup\mathcal{G}^2\ldots\cup\mathcal{G}^{T_2}\}$. Our proposed IRCD is summarized in Algorithm \ref{alg2}.
\begin{algorithm}[htbp]
\footnotesize
\caption{Iterative Reduced-space Check Detector}
\label{alg2}
\begin{algorithmic}[1]
    \Require
      $\tilde{\pmb{y}}$, $\pmb{C}$, $\mathcal{Q}$ and $\gamma_s$.
      \State \textbf{Preparation}: Set the maximum number of iteration $T_2$, and $\mathcal{G}=\emptyset$.
    \State $// \text{Reliability Sorting:}$    
    \State Employ LMMSE detection as
    \State $\tilde{\pmb{K}}=\left(\pmb{C}^H\pmb{C}+\frac{1}{\gamma_s}\pmb{I}_{QM_d}\right)^{-1}\pmb{C}^H\tilde{\pmb{y}}$.
    \State Compute the index reliability metrics based on DAPs $\mathcal{Q}_c$ as
    \State $\rho_c = \sum_{m_d=0}^{M_d-1}\left|\tilde{K}(i^c_{m_d})\right|^2$ for $i^c_{m_d}\in\{0,\ldots,QM_d-1\}$ and $c=1,\ldots,C$.
    \State Obtain the measurements of the index reliabilities as
    \State $\mathcal{R}=\{{i_1},\ldots,{i_C}\}\quad\text{subject to }{\rho}_{i_1}\geqslant\ldots\geqslant{\rho}_{i_C}$.
    \State $// \text{Reduced-space Check Detection:}$  
    \For{$t=1$ to $T_2$}
    \State Collect the DAP $\mathcal{L}^t$ according to $\rho_{i_t}$.
    \State Carry out APM symbol estimation $\pmb{K}_d^t$ based on \eqref{Eq42} and \eqref{Eq43}.
    \State Obtain the detected results as ${\mathcal{G}}^t=\{\mathcal{I}^t,\pmb{K}_d^t\}$.
    \State Calculate the residual error as
    \State $\epsilon^t=\left\|\tilde{\pmb{y}}-\pmb{C}_{\mathcal{I}^t}\pmb{K}_d^t\right\|^2$.
    \State Group the detection candidate sets as $\mathcal{G}=\mathcal{G}\cup\mathcal{G}^t$.
    \EndFor
    \State Compute the final detection results as
    \State $(\mathcal{I}^{\text{IRCD}},\pmb{K}_d^{\text{IRCD}})=\underset{(\mathcal{I}^t,\pmb{K}_d^t)\subset\mathcal{G}}\argmin\left\|\tilde{\pmb{y}}-\pmb{C}_{\mathcal{I}^t}\pmb{K}_d^t\right\|^2$.
\State \textbf{return} $\mathcal{I}^{\text{IRCD}}$ and $\pmb{K}_d^{\text{IRCD}}$.
\end{algorithmic}
\end{algorithm}
\vspace{-1em}
\subsection{Detection Complexity Analysis}\label{section3-4}
It can be concluded from \eqref{Eq27} that the MLD evaluates all the $M_d$ STSK blocks. Hence the complexity of the optimum MLD is on the order of $\mathcal{O}[(VQ)^{M_d}]$, which is excessive for high values of $M_d$.

Based on our analysis in Section \ref{section3-2}, it can be readily shown that the complexity of the PRCGD relies on the number of iterations and the value of $C_t$. In more detail, we have the best scenario, if the PRCGD terminates after the first iteration and only a single DAP is considered. Therefore, the corresponding complexity is given by $\mathcal{O}(M_dV)$, since only the symbol-wise detection of \eqref{Eq43} is employed. By contrast, the worst-case scenario is when all the DAPs in $\mathcal{Q}$ are tested. Since there are $C=Q^{M_d}$ DAPs, the overall complexity is on the order of $\mathcal{O}(Q^{M_d}M_dV)$. In more general cases, our PRCGD only has to consider a subset of $\mathcal{Q}$, having $C_1<C$ DAPs. Under this condition, the complexity of the PRCGD is on the order of $\mathcal{O}(C_1M_dV)$.

Based on Section \ref{section3-3}, the complexity of each IRCD iteration is on the order of $\mathcal{O}({M_d}V)$. Hence, the overall complexity of IRCD is given by $\mathcal{O}(T_2{M_d}V)$, and the worst scenario happens when the IRCD terminates after $T_2=Q^{M_d}$ iterations, i.e., all the DAPs are tested. However, since the IRCD measures the index reliabilities of all DAPs, near-ML detection performance can be achieved by only testing a small subset of the entire DAP set, which is shown in Section \ref{Section5}. Therefore, it can be readily demonstrated that the complexity of our IRCD can be significantly lower than that of the MLD.
\vspace{-1.5em}
\subsection{System Complexity Analysis}\label{section3-5}
{The system complexity of the SM/STSK-based schemes is identical to the corresponding MLD complexity \cite{5599264}, hence we provide our system complexity analysis in this section. We commence by specifying the SIMO-OTFS, SM-OTFS and STSK-OFDM-MA systems with the aid of their parameters as $(N_r,V)$, $(N_t,N_r,V)$ and $(N_t,N_r,T_c,Q,V)$, respectively.} {Since all the combinations of the symbols have to be evaluated in the MLD, the complexity of SIMO-OTFS $(N_r,V)$ and SM-OTFS $(N_t,N_r,V)$ is on the order of $\mathcal{O}(V^{MN})$ and $\mathcal{O}[(N_tV)^{MN}]$ \cite{10129061}, respectively.}

{The STSK-aided OFDM-MA (STSK-OFDM-MA) $(N_t,N_r,T_c,Q,V)$ system can be viewed as a special case of STSK-OTFS-MA $(N_t,N_r,T_c,Q,V)$ associated with $N=1$. Therefore, the corresponding system complexity can be expressed as $\mathcal{O}[(QV)^M]$.}
\vspace{-1em}
\section{Performance Analysis of the Single-User System and Dispersion Matrix Design}\label{Section4}
In this section, we commence with the BER analysis of the single-user STSK-OTFS-MA system and derive its performance upper-bound, referred to as the SU-UPEP. Then the diversity order and ST coding gain are derived. Moreover, the single-user DCMC capacity of the STSK-OTFS-MA system is discussed. Finally, based on the SU-UPEP and the DCMC capacity, we gain deeper insights into the criteria of DM design in Section \ref{section4-4}, and the algorithm for optimizing the DMs is derived.
\subsection{Analysis of Single-User Bit Error Ratio Performance}\label{section4-1}
The vector-form input-output relationship in \eqref{Eq17} can be expressed as $\left(\pmb{y}^{(u)}_{n_r,n_t,t_c}\right)^T=\breve{\pmb{h}}^{(u)}_{n_r,n_t}\breve{\pmb{X}}^{(u)}_{n_t,t_c}+\left(\pmb{n}^{(u)}_{n_r,n_t,t_c}\right)^T$, where $\breve{\pmb{h}}^{(u)}_{n_r,n_t}=\left[\breve{h}^{(u)}_{n_r,n_t}(1),\ldots,\breve{h}^{(u)}_{n_r,n_t}(P)\right]$ with $\breve{h}^{(u)}_{n_r,n_t}(i)=h^{(u)}_{i,n_r,n_t}e^{-j2\pi\nu_i\tau_i},\forall i$, and the $m_d$th column of $\breve{\pmb{X}}^{(u)}_{n_t,t_c}\in\mathbb{C}^{P\times M_d}$ can be obtained as
\begin{align}\label{Eq4-2}
\breve{\pmb{X}}^{(u)}_{n_t,t_c}[:,m_d]=\begin{bmatrix}
		x^{(u)}_{n_t,t_c}([k-k_1]_N+N[l-l_1]_M)\\
		\vdots\\
		x^{(u)}_{n_t,t_c}([k-k_P]_N+N[l-l_P]_M)
			\end{bmatrix},
\end{align}
where we have $m_d=k+Nl$ for $k=0,\ldots,N-1$ and $l=0,\ldots,M-1$. Similar to \eqref{Eq19}, the $n_r$th received codeword vector within the $t_c$th OTFS time-slot can be formulated as
\begin{align}\label{Eq4-3}	
\pmb{y}^T_{n_r,t_c}&=\sum_{u=0}^{U-1}\sum_{n_t=0}^{N_t-1}\breve{\pmb{h}}^{(u)}_{n_r,n_t}\breve{\pmb{X}}^{(u)}_{n_t,t_c}+\pmb{n}^T_{n_r,t_c}=\sum_{u=0}^{U-1}\breve{\pmb{h}}^{(u)}_{n_r}\breve{\pmb{X}}^{(u)}_{t_c}+\pmb{n}^T_{n_r,t_c},
	\end{align}
where we have $\breve{\pmb{h}}^{(u)}_{n_r}=\left[\breve{\pmb{h}}^{(u)}_{0,n_r},\ldots,\breve{\pmb{h}}^{(u)}_{N_t-1,n_r}\right]$ and $\breve{\pmb{X}}^{(u)}_{t_c}=\text{sta}\{\breve{\pmb{X}}^{(u)}_{n_t,t_c}\}|_{n_t=0}^{N_t-1}$. Therefore, the end-to-end input-output relationship of the $t_c$th time-slot is given by $\breve{\pmb{Y}}_{t_c}=\sum_{u=0}^{U-1}\breve{\pmb{H}}^{(u)}\breve{\pmb{X}}^{(u)}_{t_c}+\breve{\pmb{n}}_{t_c}$ for $t_c=0,\ldots,T_c-1$, where $\breve{\pmb{Y}}_{t_c}=\left[\pmb{y}_{0,t_c},\ldots,\pmb{y}_{N_r-1,t_c}\right]^T$ is the matrix of received signal, $\breve{\pmb{n}}_{t_c}=\left[\pmb{n}_{0,t_c},\ldots,\pmb{n}_{N_r-1,t_c}\right]^T$ denotes the noise matrix, and the channel matrix can be expressed as $\breve{\pmb{H}}^{(u)}=\text{sta}\{\breve{\pmb{h}}^{(u)}_{n_r}\}|_{n_r}^{N_r-1}$. Finally, by defining $\pmb{Y}=\left[\breve{\pmb{Y}}_{0},\ldots,\breve{\pmb{Y}}_{T_c-1}\right]$, $\breve{\pmb{X}}^{(u)}=\left[\breve{\pmb{X}}^{(u)}_{0},\ldots,\breve{\pmb{X}}^{(u)}_{T_c-1}\right]$ and $\breve{\pmb{n}}=\left[\breve{\pmb{n}}_{0},\ldots,\breve{\pmb{n}}_{T_c-1}\right]$, the input-output relationship for the entire transmitted frame can be formulated as
\begin{align}\label{Eq4-5}
	\pmb{Y}&=\sum_{u=0}^{U-1}\breve{\pmb{H}}^{(u)}\breve{\pmb{X}}^{(u)}+\breve{\pmb{n}}=\pmb{H}\pmb{X}+\breve{\pmb{n}},
		\end{align}
where we have $\pmb{X}=\text{sta}\{\breve{\pmb{X}}^{(u)}\}|_{u=0}^{U-1}$ and $\pmb{H}=\left[\breve{\pmb{H}}^{(u)},\ldots,\breve{\pmb{H}}^{(U-1)}\right]$. In a single-user scenario, we have $\pmb{H}\in\mathbb{C}^{N_r\times PN_t}$ and $\pmb{X}\in\mathbb{C}^{PN_t\times M_dT_c}$. For notational simplicity, the index $(u)$ is omitted in the rest of this section, since only a single user is considered. Therefore, the MLD associated with the input-output relationship shown in \eqref{Eq4-5} can be formulated as $\pmb{X}^{\text{ML}}=\underset{\pmb{D}_i\in\mathcal{D}}{\arg\min} \left\{||\pmb{Y}-\pmb{H}\pmb{D}_i||^2\right\}$, where the equivalent candidate space $\mathcal{D}$ can be formulated based on $\mathcal{B}$ in \eqref{Eq30} and the mapping relationship in \eqref{Eq4-2}, yielding $\mathcal{D}=\left\{\pmb{D}_1,\ldots,\pmb{D}_{2^L}: \pmb{D}_i\in\mathbb{C}^{PN_t\times M_dT_c},i=1,\ldots,2^{L}\right\}$.

Let us consider the pairwise error event $\{\pmb{X}^c\rightarrow \pmb{X}^e\}$, where $\pmb{X}^c=\pmb{D}_i$ denotes the transmitted codeword matrix, while $\pmb{X}^e=\pmb{D}_j$, $\forall i\neq j$, represents the erroneous detection results of the MLD, i.e., we have $\pmb{D}_i\neq\pmb{D}_j$ for $\pmb{D}_i, \pmb{D}_j\in\mathcal{D}$. Let us define furthermore the error matrix space $\mathcal{\mathcal{E}}=\{\pmb{E}=\pmb{D}_i-\pmb{D}_j,\forall \pmb{D}_i,\pmb{D}_j\in\mathcal{D},\forall i\neq j\}$. Then, the conditional PEP for a given channel matrix $\pmb{H}$ is obtained as $P_E(\pmb{X}^c\rightarrow\pmb{X}^e|\pmb{H})=P\left(||\pmb{Y}-\pmb{H}\pmb{X}^c||^2\geqslant||\pmb{Y}-\pmb{H}\pmb{X}^e||^2\right)$. Let us denote the elements of the corresponding matrices as $H(a,b)$, $X(b,c)$, $n(a,c)$ and $Y(a,c)$ for $a=0,\ldots,N_r-1$, $b=0,\ldots,M_dN_t-1$ and $c=0,\ldots,M_dT_c-1$, respectively. Then, after some further algebraic simplifications, it can be shown that $P_E(\pmb{X}^c\rightarrow\pmb{X}^e|\pmb{H})$ may be written as
\begin{align}\label{Eq4-7}
P_E(\pmb{X}^c\rightarrow\pmb{X}^e|\pmb{H})=&P\left(\sum_{c=0}^{M_dT_c-1}\sum_{a=0}^{N_r-1}\Re\left\{n^*(a,c)\sum_{b=0}^{PN_t-1}z\right\}\right.\nonumber\\
\geqslant &\left.\frac{1}{2}\sum_{c=0}^{M_dT_c-1}\sum_{a=0}^{N_r-1}\left|\sum_{b=0}^{PN_t-1}z\right|^2\right),
\end{align}
where $z=H(a,b)\left[X^{e}(b,c)-X^{c}(b,c)\right]$. Consequently, when defining the modified Euclidean distance between two codeword matrices $\pmb{X}^e$ and $\pmb{X}^c$ as $\delta(\pmb{X}^c,\pmb{X}^e)=\sum_{c=0}^{M_dT_c-1}\sum_{a=0}^{N_r-1}\left|\sum_{b=0}^{PN_t-1}z\right|^2=||\pmb{H}(\pmb{X}^e-\pmb{X}^c)||^2$, and considering that $\sum_{c=0}^{M_dT_c-1}\sum_{a=0}^{N_r-1}\Re\left\{n^*(a,c)\sum_{b=0}^{PN_t-1}z\right\}$ in \eqref{Eq4-7} is a complex-valued Gaussian random variable with zero mean and a variance of $||\pmb{H}(\pmb{X}^e-\pmb{X}^c)||^2/2\gamma$, \eqref{Eq4-7} can be further expressed as
\begin{align}\label{Eq4-9}
	P_E(\pmb{X}^c\rightarrow\pmb{X}^e|\pmb{H})=Q\left(\sqrt{\frac{\gamma\delta(\pmb{X}^c,\pmb{X}^e)}{2}}\right),
\end{align}
where $Q(x)=\frac{1}{\pi}\int_{0}^{\pi/2} \exp\left(-\frac{x^2}{2\sin^2\theta}d\theta\right)$ is the Gaussian $Q$-function with $x>0$. Note that \eqref{Eq4-9} can be alternatively represented as $P_E(\pmb{X}^c\rightarrow\pmb{X}^e|\pmb{H})=\frac{1}{\pi}\int_0^{\frac{\pi}{2}}\exp\left(-\frac{\gamma\delta(\pmb{X}^c,\pmb{X}^e)}{4\sin^2\theta}\right)d\theta$ \cite{yang2009multicarrier}. By averaging the integration of $P_E(\pmb{X}^c\rightarrow\pmb{X}^e|\pmb{H})$ with respect to the distribution of $\delta(\pmb{X}^c,\pmb{X}^e)$, the SU-UPEP can be formulated as \cite{yang2009multicarrier}
\begin{align}\label{Eq4-11}
	P_E(\pmb{X}^c\rightarrow\pmb{X}^e)&=\mathbb{E}_{\pmb{H}}\left[\frac{1}{\pi}\int_0^{\frac{\pi}{2}}\exp\left(-\frac{\gamma\delta(\pmb{X}^c,\pmb{X}^e)}{4\sin^2\theta}\right)d\theta\right]\nonumber\\
	&=\frac{1}{\pi}\int_0^{\frac{\pi}{2}}\Gamma_{\delta(\pmb{X}^c,\pmb{X}^e)}\left(-\frac{\gamma}{4\sin^2\theta}\right)d\theta,
		\end{align}
where $\Gamma_{\delta(\pmb{X}^c,\pmb{X}^e)}(t)$ is the moment generating function (MGF) of $\delta(\pmb{X}^c,\pmb{X}^e)$. Moreover, based on the analysis in \cite{979326}, $\delta(\pmb{X}^c,\pmb{X}^e)$ can be formulated as $\delta(\pmb{X}^c,\pmb{X}^e)=\sum_{i=0}^{M_dT_c-1}||\pmb{H}\left(\pmb{X}^c[:,i]-\pmb{X}^e[:,i]\right)||^2\\=\tilde{\pmb{h}}[\pmb{I}_{N_r}\otimes\pmb{R}]\tilde{\pmb{h}}^H$, where the codeword difference matrix is defined as $\pmb{R}=(\pmb{X}^e-\pmb{X}^c)(\pmb{X}^e-\pmb{X}^c)^H$ and $\tilde{\pmb{h}}=(\text{vec}(\pmb{H}^T))^T\in\mathbb{C}^{1\times PN_rN_t}$. Since the non-zero elements of $\tilde{\pmb{h}}$ obeys zero mean and variance of $1/2P$ per real dimension Gaussian distribution \cite{8686339}, we can derive the MGF $\Gamma_{\delta(\pmb{X}^c,\pmb{X}^e)}(t)$ based on the approach of \cite{979326}, yielding
\begin{align}\label{Eq4-13}
	\Gamma_{\delta(\pmb{X}^c,\pmb{X}^e)}(t)=\det[\pmb{I}_{PN_rN_t}-t(\pmb{I}_{N_r}\otimes\pmb{R})/P]^{-1}.
	\end{align}
Upon substituting \eqref{Eq4-13} into \eqref{Eq4-11}, the SU-UPEP can now be expressed as
\begin{align}\label{Eq4-14}
	P_E(\pmb{X}^c\rightarrow\pmb{X}^e)&=\frac{1}{\pi}\int_0^{\frac{\pi}{2}}\left[\det\left(\pmb{I}_{N_rPN_t}+\frac{\gamma}{4P\sin^2\theta}(\pmb{I}_{N_r}\otimes\pmb{R})\right)\right]^{-1}d\theta\nonumber\\
	&=\frac{1}{\pi}\int_0^{\frac{\pi}{2}}\left[\det\left(\pmb{I}_{PN_t}+\frac{\gamma}{4P\sin^2\theta}\pmb{R}\right)\right]^{-N_r}d\theta.
\end{align}

Let us define $r=\text{rank}(\pmb{R})$ and the nonzero eigenvalues of $\pmb{R}$ as $\{\lambda_1,\ldots,\lambda_r\}$. Then \eqref{Eq4-14} can be expressed as
\begin{align}\label{Eq4-15}
	P_E(\pmb{X}^c\rightarrow\pmb{X}^e)&=\frac{1}{\pi}\int_0^{\frac{\pi}{2}}\left[\prod_{j=1}^{r}\left(1+\frac{\lambda_j\gamma}{4P\sin^2\theta}\right)\right]^{-N_r}d\theta.	
\end{align}

Finally, by leveraging the union bound technique, the average bit error ratio (ABER) of the single-user STSK-OTFS-MA system can be approximated as
\begin{align}\label{Eq4-16}
P_e\approx\frac{1}{2^{L}L}\sum_{\pmb{b}^c}\sum_{\pmb{b}^e}D_b(\pmb{b}^c,\pmb{b}^e)P_E(\pmb{X}^c\rightarrow\pmb{X}^e),
\end{align}
where $D_b(\cdot,\cdot)$ denotes the Hamming distance function between two bit sequences, while $\pmb{b}^c$ and $\pmb{b}^e$ are the corresponding binary representations of $\pmb{X}^c$ and $\pmb{X}^e$.
\subsection{Diversity Order and Coding Gain}\label{section4-2}
In \eqref{Eq4-15} we have ${\lambda_j\gamma}/{(4P\sin^2\theta)}\geqslant{\lambda_j\gamma}/{4P}$. Hence, the upper-bound of $P_E(\pmb{X}^c\rightarrow\pmb{X}^e)$ can be formulated as
\begin{align}\label{Eq4-18}
	P_E(\pmb{X}^c\rightarrow\pmb{X}^e)\leqslant\frac{1}{2}\left[\prod_{j=1}^{r}\left(1+\frac{\lambda_j\gamma}{4P}\right)\right]^{-N_r}.
\end{align}
Moreover, for high SNRs ($\gamma\gg 1$), \eqref{Eq4-18} can be formulated as
\begin{align}\label{Eq4-19}
	P_E(\pmb{X}^c\rightarrow\pmb{X}^e)\leqslant\frac{1}{2}\left[\left(\prod_{j=1}^{r}\lambda_j\right)^{1/r}\left(\frac{\gamma}{4P}\right)\right]^{-r N_r},	
	\end{align}
where the exponent of the SNR is often referred to as the \textbf{diversity order} obtained by the MLD, which is
\begin{align}\label{Eq4-20}
	G_D=\underset{\forall \pmb{E}\in\mathcal{E}}\min rN_r=\underset{\forall \pmb{E}\in\mathcal{E}}\min\text{rank}(\pmb{R})\cdot N_r,
\end{align}
and the maximum achievable diversity order is $G_{\text{D,max}}=\min\{PN_t,M_dT_c\}\cdot N_r$. Given the values of $P$ and $M_d$, it can be observed that the maximum achievable diversity order depends on the settings of $N_t$ and $T_c$. Although the diversity order can be increased by increasing the value of $T_c$ in the case of $(PN_t>M_dT_c)$, the transmit diversity order cannot be further improved, if we increase $M_dT_c$ beyond $PN_t$. In more detail, the system with a lower value of $T_c$ may result in a higher transmission rate in \eqref{Eq30} as well as a low computation complexity, since the dimension of DMs is lower.

The {coding gain} of the STSK-OTFS-MA system can be expressed as $G_C=\underset{\forall \pmb{E}\in\mathcal{E}}\min\left(\prod_{j=1}^{r}\lambda_j\right)^{1/r}$. It can be inferred from \eqref{Eq4-19} that the diversity order $G_D$ dominates the decay-rate of the SU-UPEP as the value of SNR increases. Furthermore, the coding gains $G_C$ determines the horizontal shift of the STSK-aided SU-UPEP curve from the benchmark SU-UPEP curve of $\frac{1}{2}(\gamma/4P)^{-G_D}$.
\vspace{-1.5em}
\subsection{DCMC Capacity}\label{section4-3}
Now we derive the DCMC capacity of the single-user STSK-OTFS-MA scheme. Based on \eqref{Eq27} and \eqref{Eq30}, the single-user DCMC capacity can be formulated as \cite{1608632}
\begin{align}\label{Eq4-27}	
C_{\text{DCMC}}&=\frac{1}{M_d T_c}\underset{p(\pmb{B}_1)\ldots p(\pmb{B}_{2^L})}\max\sum_{i=1}^{2^L}\int_{-\infty}^{\infty}\ldots\int_{-\infty}^{\infty}p(\tilde{\pmb{y}}|\pmb{B}_i)p(\pmb{B}_i)\nonumber\\
	&\times\log_2\left[\frac{p(\tilde{\pmb{y}}|\pmb{B}_i)}{\sum_{j=1}^{2^L}p(\tilde{\pmb{y}}|\pmb{B}_j)p(\pmb{B}_j)}\right]d\tilde{\pmb{y}}.
	\end{align}
where $p(\tilde{\pmb{y}}|\pmb{B}_i)$ is given by \eqref{Eq31}, when assuming that $\pmb{B}_i$ is transmitted. It should be noted that \eqref{Eq4-27} is maximized under the condition that all candidate matrices in the space $\mathcal{B}$ are independent and equiprobable, i.e., $p(\pmb{B}_i)=1/2^L,\forall i$. Hence we have
\begin{align}\label{Eq4-28}
	\log_2\left[\frac{p(\tilde{\pmb{y}}|\pmb{B}_i)}{\sum\limits_{j=1}^{2^L}p(\tilde{\pmb{y}}|\pmb{B}_j)p(\pmb{B}_j)}\right]=\log_2\left(2^L\right)-\log_2\sum_{j=1}^{2^L}\exp(\Psi_{i,j}),
	\end{align}
where $\Psi_{i,j}=\gamma\left[-||\pmb{C}(\pmb{B}_i-\pmb{B}_j)+\tilde{\pmb{n}}||^2+||\tilde{\pmb{n}}||^2\right]$ can be formulated by substituting \eqref{Eq31} into \eqref{Eq4-28}. Therefore, based on the assumption that the codeword matrix candidates are transmitted at the same probabilities, the DCMC capacity of the single-user STSK-OTFS-MA system can be expressed as
\begin{align}\label{Eq4-30}
	C_{\text{DCMC}}&=\frac{1}{M_d T_c}\left\{L-\frac{1}{2^L}\sum_{i=1}^{2^L}\mathbb{E}_{\pmb{C}}\left[\log_2\sum_{j=1}^{2^L}\exp\left(\Psi_{i,j}\right)\right]\right\}.
\end{align}
\vspace{-2.5em}
\subsection{Design of Dispersion Matrices}\label{section4-4}
Based on the above analysis, let us now discuss the design of DMs. To obtain the best system performance, the DMs can be designed based either on the SU-UPEP of \eqref{Eq4-11} or on the DCMC capacity of \eqref{Eq4-30}.

{The asymptotic diversity order of the OTFS system is one at an infinite SNR, and higher diversity can be attained in finite-SNR scenarios \cite{8686339}. However, similar to OFDM, OTFS also requires transmit precoding schemes to attain full diversity \cite{8686339}. Therefore, the DMs of both STSK-OTFS-MA and STSK-OFDM-MA systems can be designed with the aid of our proposed Algorithm \ref{alg3}.}

It can be shown that the above two methods lead to the same design, detailed as follows.

\emph{Proposition 1:} The design of DMs aiming for minimizing the SU-UPEP in the high-SNR region of \eqref{Eq4-19} and that aiming for maximizing the DCMC capacity of \eqref{Eq4-30} lead to the same results, which can be formulated as \textbf{maximizing} the following metrics:
\begin{align}\label{Eq4-31}
\Lambda_{\text{D}}=\underset{\forall \pmb{E}\in\mathcal{E}}\min\text{rank}(\pmb{R}),\quad \Lambda_{\text{C}}=\underset{\forall \pmb{E}\in\mathcal{E}}\min\prod_{j=1}^{r}\lambda_j.
\end{align}
\emph{Proof:} It can be readily shown that the elements of $\{\pmb{Y},\breve{\pmb{n}}\}$ in \eqref{Eq4-5} are the interleaved versions of the elements of $\{\tilde{\pmb{y}},\tilde{\pmb{n}}\}$ in \eqref{Eq27}. Therefore, given a pairwise error event $\{\pmb{X}^c\rightarrow \pmb{X}^e\}$ and a SNR $\gamma$, we have $||\pmb{C}(\pmb{B}_i-\pmb{B}_j)||^2=||\pmb{H}(\pmb{D}_i-\pmb{D}_j)||^2$. Furthermore, similar to the derivation shown in Section \ref{section4-1}, it can be readily shown that based on \eqref{Eq4-30} and the MGF technique, maximizing the DCMC capacity is equivalent to minimizing
\begin{align}\label{Eq4-32}	
\mathbb{E}_{\pmb{H}}\left[-\gamma\delta(\pmb{X}^c,\pmb{X}^e)\right]\leqslant\frac{1}{2}\left[\prod_{j=1}^{r}\left(1+\frac{\lambda_j\gamma}{P}\right)\right]^{-N_r},
\end{align}
which can be further upper-bounded by $\frac{1}{2}\left[\left(\prod_{j=1}^{r}\lambda_j\right)^{1/r}\left(\frac{\gamma}{P}\right)\right]^{-r N_r}$ under the condition of $\gamma\gg 1$. It may now be observed that \eqref{Eq4-19} and \eqref{Eq4-32} are in similar forms. Note that it is more important to maximize the diversity order $G_D$ in \eqref{Eq4-20} than to maximize the coding gain $G_C$, since it is the diversity order that dominates the slope of the ABER curve \cite{8478829}. Since the value of $N_r$ is fixed, we arrive at the criteria $\Lambda_{\text{D}}$ in \eqref{Eq4-31} by searching for the minimum rank value within the corresponding error matrix space $\mathcal{E}$. Furthermore, the coding gain is given by $\underset{\forall \pmb{E}\in\mathcal{E}}\min\left(\prod_{j=1}^{r}\lambda_j\right)^{\frac{1}{\Lambda_{\text{D,max}}}}$. Since $\Lambda_{\text{D,max}}$ is a constant for $\forall \pmb{E}\in\mathcal{E}$, the corresponding criteria $\Lambda_{\text{C}}$ can be derived in \eqref{Eq4-31}. \hfill $\blacksquare$

Based on the above analysis, let us now delve into the details of designing the DM set. Let us assume that $\tilde{N}$ consecutive Monte Carlo simulations are employed. Let us define the DM set, design criteria, codeword difference matrix, error matrix, and error matrix space for the $\tilde{n}$th experiment as $\mathcal{A}_{\tilde{n}}$, $\Lambda_{\text{D},\tilde{n}}$, $\Lambda_{\text{C},\tilde{n}}$, $\pmb{R}_{\tilde{n}}$, $\pmb{E}_{\tilde{n}}$ and $\mathcal{E}_{\tilde{n}}$ for $\tilde{n}=1,\ldots,\tilde{N}$, respectively. For the $\tilde{n}$th experiment, we first randomly generate the $Q$ full-rank $(\bar{T}\times\bar{T})$-dimensional unitary matrices $\tilde{\pmb{A}}_{q,\tilde{n}}$ for $q=1,\ldots,Q$, where $\bar{T}=\max\{N_t,T_c\}$. Then the DMs can be formulated as
\begin{align}\label{Eq4-22}
	\pmb{A}_{q,\tilde{n}}=\begin{cases}
		\tilde{\pmb{A}}_{q,\tilde{n}}[:,1:T_c], &\mbox{if }N_t>T_c\\
		\sqrt{\frac{T_c}{N_t}}\tilde{\pmb{A}}_{q,\tilde{n}}[1:N_t,:], &\mbox{if }T_c>N_t,
		\end{cases}
\end{align}
where the constant $\sqrt{\frac{T_c}{N_t}}$ is used for satisfying the power constraint in \eqref{Eq2}, and we have $\mathcal{A}_{\tilde{n}}=\{\pmb{A}_{1,\tilde{n}},\ldots,\pmb{A}_{Q,\tilde{n}}\}$. Moreover, the DM set of all $\tilde{N}$ simulations is denoted by $\mathcal{A}=\{\mathcal{A}_1,\ldots,\mathcal{A}_{\tilde{N}}\}$.  Assuming that there are $\breve{N}$ out of $\tilde{N}$ DM sets that maximize the diversity order, it can be readily shown from \eqref{Eq4-31} that the candidates of the optimal DM set are obtained as $\breve{\mathcal{A}}=\underset{\mathcal{A}_{\tilde{n}}\subset\mathcal{A}}\argmax \Lambda_{\text{D},\tilde{n}}=\underset{\mathcal{A}_{\tilde{n}}\subset\mathcal{A}}\argmax\left\{\underset{\forall\pmb{E}_{\tilde{n}}\in\mathcal{E}_{\tilde{n}}}\min\{\text{rank}(\pmb{R}_{\tilde{n}})\}\right\}$, where $\breve{\mathcal{A}}=\left\{\breve{\mathcal{A}}_{1},\ldots,\breve{\mathcal{A}}_{\breve{N}}\right\},\breve{\mathcal{A}}_{\breve{n}}\subset\mathcal{A}, \forall \breve{n}$. Finally, the optimal DM set can be obtained as $\mathcal{A}^{\text{opt}}=\underset{\breve{\mathcal{A}}_{\breve{n}}\subset\breve{\mathcal{A}}}\argmax\Lambda_{\text{C},\breve{n}}=\underset{\breve{\mathcal{A}}_{\breve{n}}\subset\breve{\mathcal{A}}}\argmax\left\{\underset{\forall\pmb{E}_{\breve{n}}\in\mathcal{\mathcal{E}}_{\breve{n}}}\min\prod_{j=1}^{r}\lambda_j\right\}$. The proposed DM design method is summarized in Algorithm \ref{alg3}.
\begin{algorithm}[h!]
\footnotesize
\caption{Dispersion Matrix Design}
\label{alg3}
\begin{algorithmic}[1]
    \Require
      The values of $Q$, $T_c$, $N_t$ and $\bar{T}=\max\{N_t,T_c\}$.
      \State \textbf{Preparation}: Set $\tilde{N}$ as the number of Monte Carlo simulations. 
    \For{$\tilde{n}=1$ to $\tilde{N}$}
    \State Randomly generate $Q$ full-rank $(\bar{T}\times\bar{T})$-dimensional
    \Statex\quad\ unitary matrices $\bar{\mathcal{A}}_{\tilde{n}}=\{\bar{\pmb{A}}_{1,\tilde{n}},\ldots,\bar{\pmb{A}}_{Q,\tilde{n}}\}$.   
    \State Generate the DM set $\mathcal{A}_{\tilde{n}}=\{\pmb{A}_{1,\tilde{n}},\ldots,\pmb{A}_{Q,\tilde{n}}\}$ 
    \Statex\quad\ based on \eqref{Eq4-22} as
    \State $\pmb{A}_{q,\tilde{n}}=\begin{cases}
		\tilde{\pmb{A}}_{q,\tilde{n}}[:,1:T_c], &\mbox{if }N_t>T_c\\
		\sqrt{\frac{T_c}{N_t}}\tilde{\pmb{A}}_{q,\tilde{n}}[1:N_t,:], &\mbox{if }T_c>N_t,
		\end{cases}$
    \EndFor
    \State Collect all DM matrices $\mathcal{A}=\{\mathcal{A}_1,\ldots,\mathcal{A}_{\tilde{N}}\}$.
    \State Obtain $\breve{N}$ out of $\tilde{N}$ DM sets that achieve the maximum diversity order as 
    \State $\breve{\mathcal{A}}=\underset{\mathcal{A}_{\tilde{n}}\subset\mathcal{A}}\argmax \Lambda_{\text{D},\tilde{n}}=\underset{\mathcal{A}_{\tilde{n}}\subset\mathcal{A}}\argmax\left\{\underset{\forall\pmb{E}_{\tilde{n}}\in\mathcal{E}_{\tilde{n}}}\min\{\text{rank}(\pmb{R}_{\tilde{n}})\}\right\}$,
    \Statex where $\breve{\mathcal{A}}=\left\{\breve{\mathcal{A}}_{1},\ldots,\breve{\mathcal{A}}_{\breve{N}}\right\},\breve{\mathcal{A}}_{\breve{n}}\subset\mathcal{A}, \forall \breve{n}$.
    \State Generate the optimal DM set as
    \State $\mathcal{A}^{\text{opt}}=\underset{\breve{\mathcal{A}}_{\breve{n}}\subset\breve{\mathcal{A}}}\argmax\Lambda_{\text{C},\breve{n}}=\underset{\breve{\mathcal{A}}_{\breve{n}}\subset\tilde{\mathcal{A}}}\argmax\left\{\underset{\forall\pmb{E}_{\breve{n}}\in\mathcal{\mathcal{E}}_{\breve{n}}}\min\prod_{j=1}^{r}\lambda_j\right\}$.
\State \textbf{return} $\mathcal{A}^{\text{opt}}$.
\end{algorithmic}
\end{algorithm}
\vspace{-1em}
\section{Performance Results}\label{Section5}
In this section, we provide the simulation results for characterizing the overall performance of STSK-OTFS-MA systems. Unless specifically defined, $N=4$ time intervals are considered for an OTFS subframe and the entire OTFS frame has $T_c=2$ subframes {while resource allocation scheme 1 is invoked}. The DMs are obtained from Algorithm \ref{alg3}. The subcarrier spacing and carrier frequency are $\Delta f=15$ kHz and $f_c=4$ GHz. The normalized maximum Doppler and delay shifts are set to $k_\text{max}=M-1$ and $l_\text{max}=N-1$ \cite{8686339}, respectively. The normalized indices of delay and Doppler associated with the $i$th path are given by $l_i\in \mathcal{U}[0,l_{\text{max}}]$ and $k_i\in \mathcal{U}[-k_{\text{max}},k_{\text{max}}]$, respectively. Furthermore, the SIMO-OTFS and SM-OTFS systems are specified as $(N_r,V)$ and $(N_t,N_r,V)$, respectively. More explicitly, the SM-OTFS can also be viewed as a single-user STSK-OTFS-MA $(N_t,N_r,1,V,Q=N_t)$ system, where $\pmb{A}_q=\pmb{I}_{N_t}[:,q]$ for $q=1,\ldots,Q$ \cite{5599264}.
\begin{figure}[htbp]
\centering
\includegraphics[width=\linewidth]{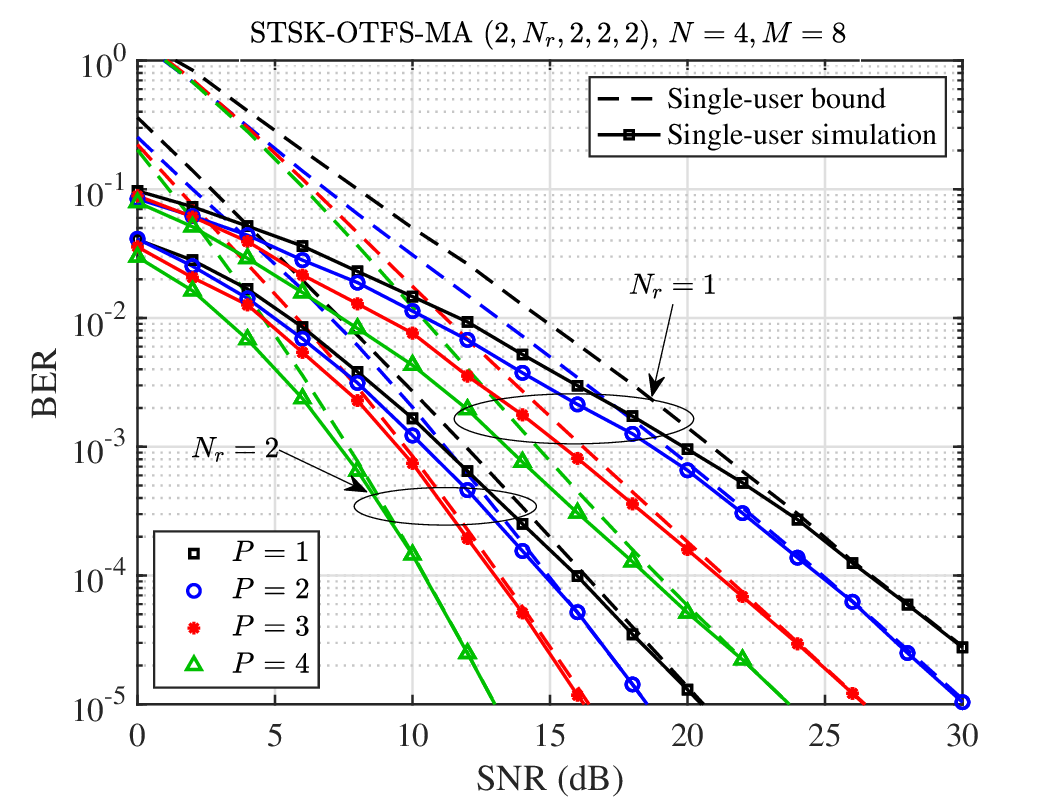}
\caption{Single-user BER performance using MLD and upper-bounds for STSK-OTFS-MA $(2,N_r,2,2,2)$ systems with $N=4$, $M=8$ and $N_r=\{1,2\}$, communicating over doubly-selective channels having the different number of paths at the same transmission rate of 1 bits/s/Hz. The upper-bounds are calculated based on \eqref{Eq4-16}.}
\label{Figure3}
\end{figure}

In Fig. \ref{Figure3}, the single-user BER performance of STSK-OTFS-MA $(2,N_r,2,2,2)$ systems detected by the MLD is compared to the corresponding BER upper-bound shown in \eqref{Eq4-16}, where $M=8$ subcarriers are employed. Explicitly, the BER performance is evaluated with different numbers of RAs and channel paths. Based on the simulation results of Fig. \ref{Figure3}, we have the following observations. Firstly, as the SNR $\gamma$ increases, the analytical BER upper-bound approaches with the simulated BER. Secondly, given $P=4$, the union-bound converges to the BER of MLD in the case of $\gamma>18$ dB for $N_r=1$, and $\gamma>8$ dB for $N_r=2$ in general. And again, the upper-bounds approach the simulated BER values in the moderate- and high-SNR scenarios for other values of $P$. Furthermore, it can be observed that all the upper-bounds are tight below $10^{-3}$ BER in general. Finally, the higher the values of $P$ and $N_r$, the higher the BER performance gain achieved by our STSK-OTFS-MA system. This is because both frequency diversity and space diversity can be achieved by the proposed STSK-OTFS-MA system, and a higher diversity order can be attained as $P$ or $N_r$ increases, which are consistent with our analytical results in \eqref{Eq4-20}.
\begin{figure}[htbp]
\centering
\includegraphics[width=\linewidth]{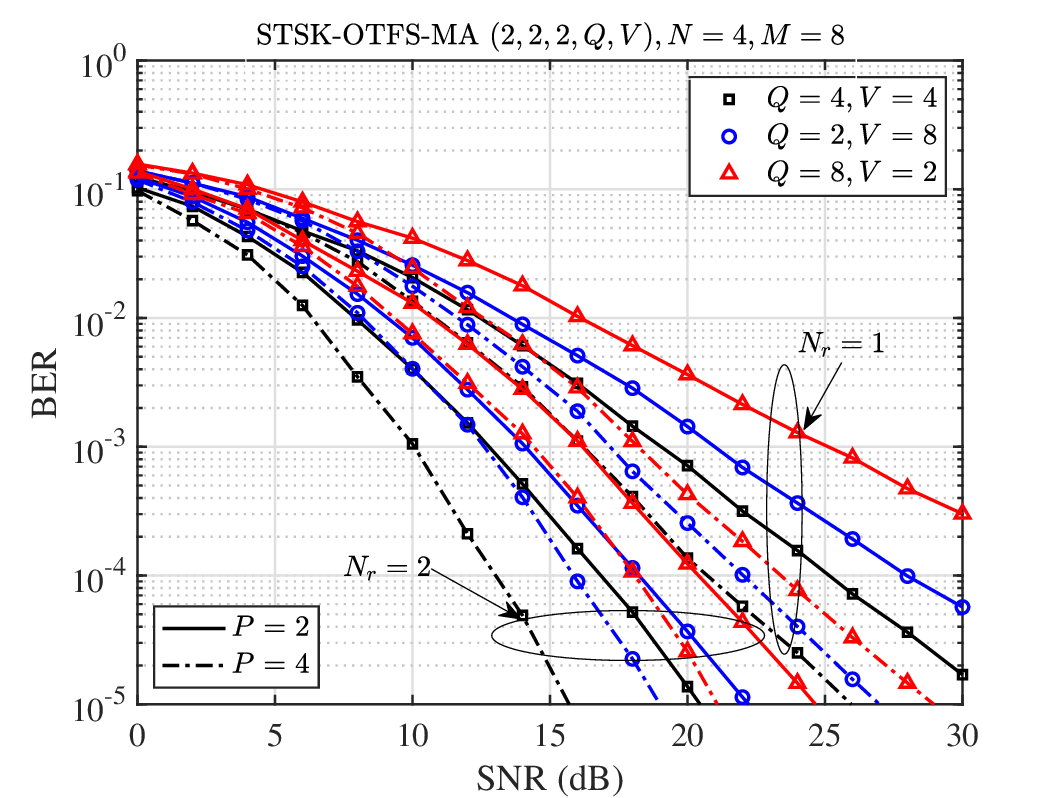}
\caption{Single-user BER performance employing MLD for STSK-OTFS-MA $(2,2,2,Q,V)$ systems with $P=\{2,4\}$ and $N_r=\{1,2\}$ but invoking a different number of DMs and modulation orders at the same transmission rate of 2 bits/s/Hz.}
\label{Figure4}
\end{figure}  

Fig. \ref{Figure4} depicts the single-user BER performance of STSK-OTFS-MA $(2,2,2,Q,V)$ systems for $P=\{2,4\}$ and $N_r=\{1,2\}$, where $M=8$ subcarriers are employed and the effect of different combinations of $\{Q,V\}$ are investigated under the constraint of $R=2$ bits/s/Hz. As shown in Fig. \ref{Figure4}, for all BER curves specified by $P=\{2,4\}$ and $N_r=\{1,2\}$, the system exploiting $Q=4$ and QPSK modulation is capable of achieving the best BER performance among the three sets of parameters $\{Q,V\}$ considered. This observation implies that for a given transmission rate $R$, an optimum combination of the parameters $\{Q,V\}$ can be found, which leads to the best BER performance. Finally, given the same other parameters used, we observe based on Fig. \ref{Figure3} and Fig. \ref{Figure4} that our STSK-OTFS-MA system is capable of attaining a better BER performance for relatively lower values of the parameters $\{Q,V\}$.
\begin{figure}[htbp]
\centering
\includegraphics[width=\linewidth]{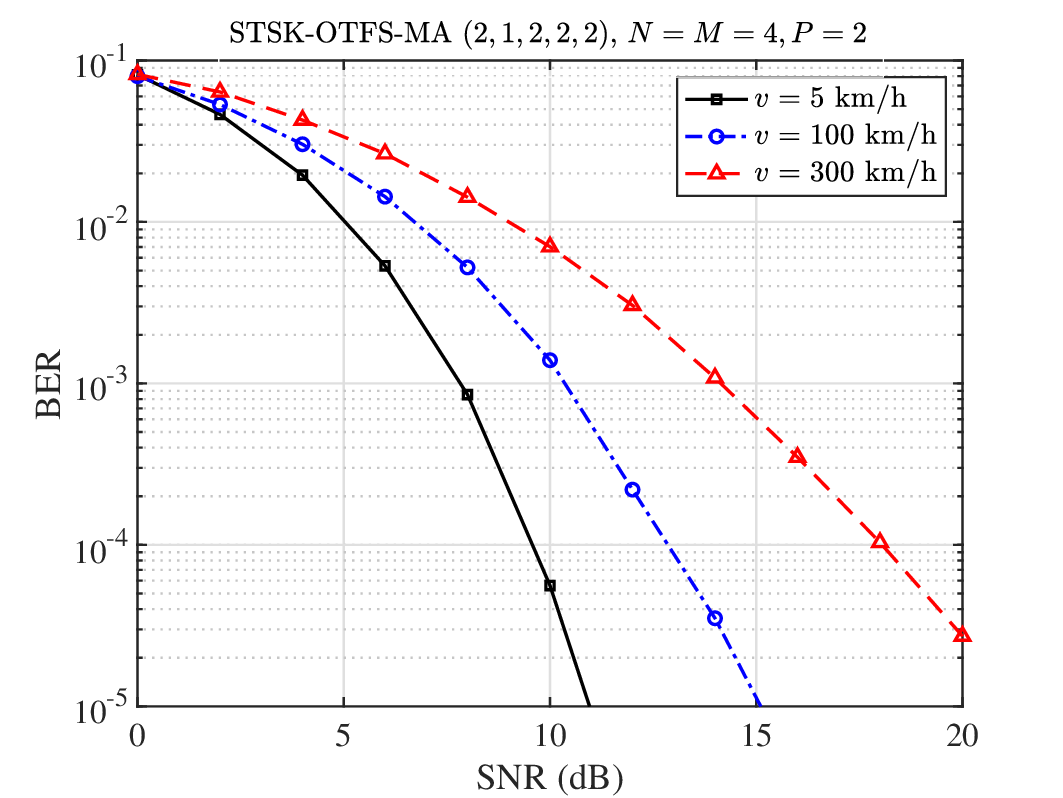}
\caption{BER performance of two-user STSK-OTFS-MA $(2,1,2,2,2)$ system using MLD with fractional delay and Doppler shifts under different mobile velocities.}
\label{Figure5-1}
\vspace{-1.5em}
\end{figure}

{In Fig. \ref{Figure5-1}, the BER performance of the two-user STSK-OTFS-MA $(2,1,2,2,2)$ system using MLD with fractional delay and Doppler shifts are investigated in the case of different mobile speeds, where we have $N=4$ and $M=4$. It is observed that the BER performance degrades as a higher speed is encountered, which is owing to the higher Doppler frequency. Explicitly, at a BER of $10^{-4}$, the $v=5$ km/h scenario yields about $2.5$ dB and $8$ dB SNR gain compared to the $v=100$ km/h and $v=300$ km/h cases.}

\begin{figure}[htbp]
\centering
\includegraphics[width=\linewidth]{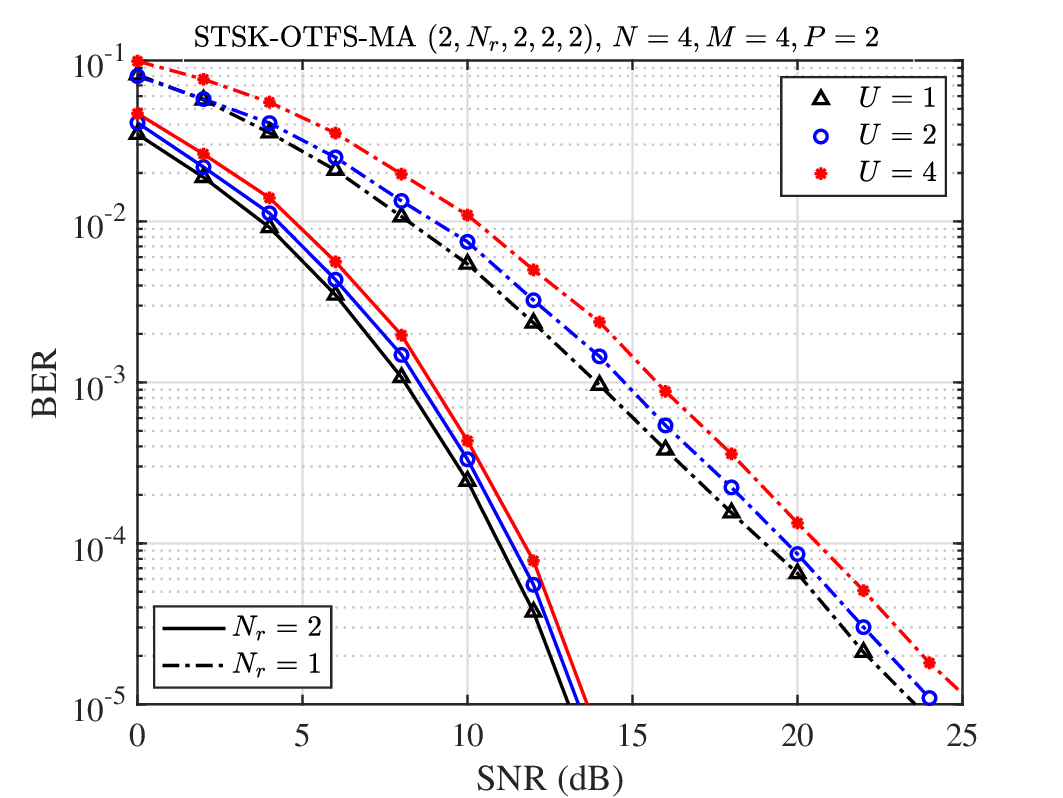}
\caption{BER performance utilizing MLD for STSK-OTFS-MA $(2,N_r,2,2,2)$  systems with $N=4$, $M=4$ and $N_r=\{1,2\}$ supporting a different number of users at the same transmission rate of $R=1$ bits/s/Hz.}
\label{Figure5}
\end{figure}

In Fig. \ref{Figure5}, we investigate the multiuser BER performance of our STSK-OTFS-MA $(2,N_r,2,2,2)$ system employing a different number of RAs. Observe from Fig. \ref{Figure5} that the BER performance degrades when the system supports more users. This is because a higher MUI is experienced as $U$ increases. More specifically, it can be observed that the four-user system equipped with $N_r=1$ RA suffers from about a 2 dB performance loss compared to the single-user system at a BER of $10^{-4}$. The corresponding performance erosion is reduced to about 1 dB for $N_r=2$. This trend is reminiscent of the channel hardening phenomenon of the classic massive MIMO uplink detection \cite{chen2018channel}.
\begin{figure}[htbp]
\centering
\includegraphics[width=\linewidth]{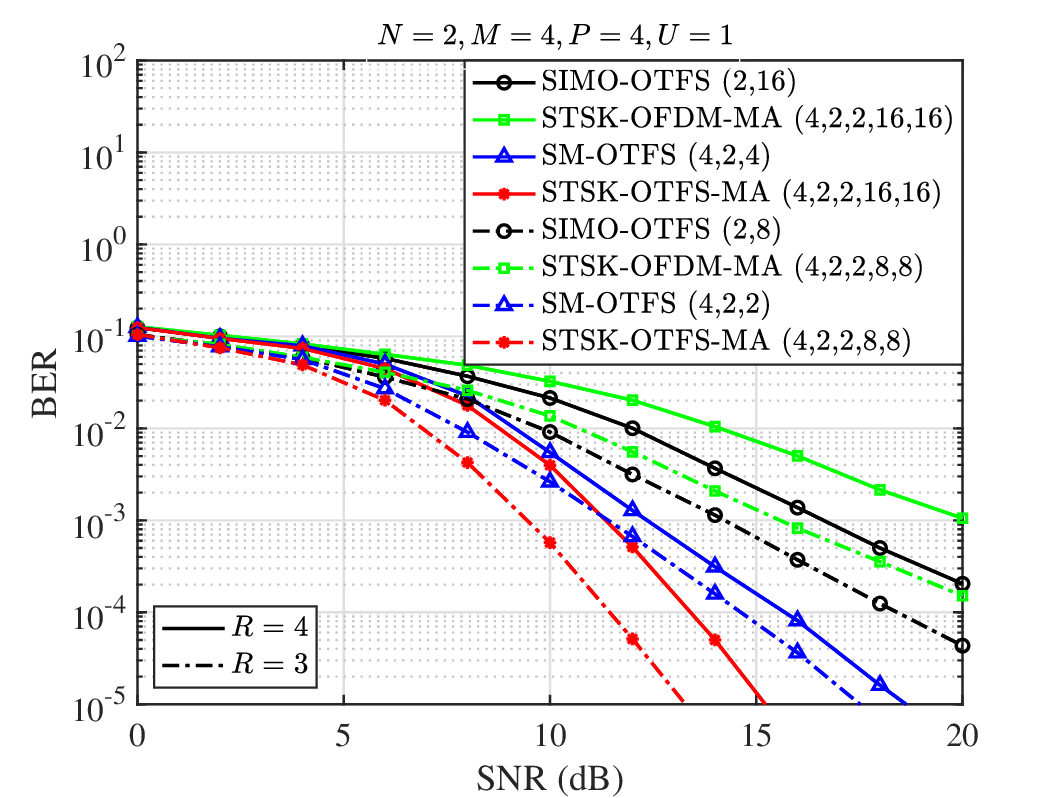}
\caption{BER performance of the conventional SIMO-OTFS scheme, the SM-OTFS scheme, the STSK-OFDM-MA scheme, and our STSK-OTFS-MA scheme both invoking MLD for the cases of the same transmission rate of $R=3$ and $R=4$ bits/s/Hz.}
\label{Figure6}
\vspace{-1em}
\end{figure}

Fig. \ref{Figure6} compares the BER performance of conventional SIMO-OTFS, of SM-OTFS as well as of the single-user {STSK-OFDM-MA} and the proposed single-user STSK-OTFS-MA systems, where $R=3$ and $R=4$ bits/s/Hz are considered, and the same values of $N_r$ are employed under the same constraint of $R$. Based on Fig. \ref{Figure6}, we have the following observations. Firstly, for a given rate, the BER performance of SM-OTFS is better than that of the conventional SIMO-OTFS. This is because SM-OTFS may rely on a lower-order modulation scheme than SIMO-OTFS. Moreover, spatial diversity can be attained by SM-OTFS. Secondly, it is found that the proposed STSK-OTFS-MA schemes are capable of achieving better performance than the SM-OTFS systems in all the scenarios considered, resulting in about 4 dB gain at a BER of $10^{-5}$ for the rate of $R=4$ bits/s/Hz, and about a 5 dB gain at a BER of $10^{-5}$ in the $R=3$ bits/s/Hz scenario. This observation can be explained by our analytical results of Section \ref{section4-2}. As an ST coding scheme, the proposed STSK-OTFS-MA system can also achieve time diversity in addition to space diversity and frequency diversity, yielding a higher diversity order than the other schemes. Additionally, the maximum coding gain can be achieved by taking full advantage of Algorithm \ref{alg3}. Therefore, our OTFS-STSK-MA scheme outperforms the SIMO-OTFS and SM-OTFS schemes. {Furthermore, STSK-OFDM-MA attains the worst BER performance at a given rate, since the ICI introduced by high-mobility channels is ignored in STSK-OFDM-MA. Specifically, at a BER of $10^{-3}$, our proposed STSK-OTFS-MA is capable of attaining about $6$ dB and $9$ dB SNR gains in the $R=3$ and $R=4$ bits/s/Hz scenarios, respectively.} Finally, the BER performances of all systems improve, as the rate is reduced due to the lower values of $Q$ and $V$ used. This observation is consistent with the conclusions in \cite{5599264}.
\begin{figure}[htbp]
\centering
\includegraphics[width=\linewidth]{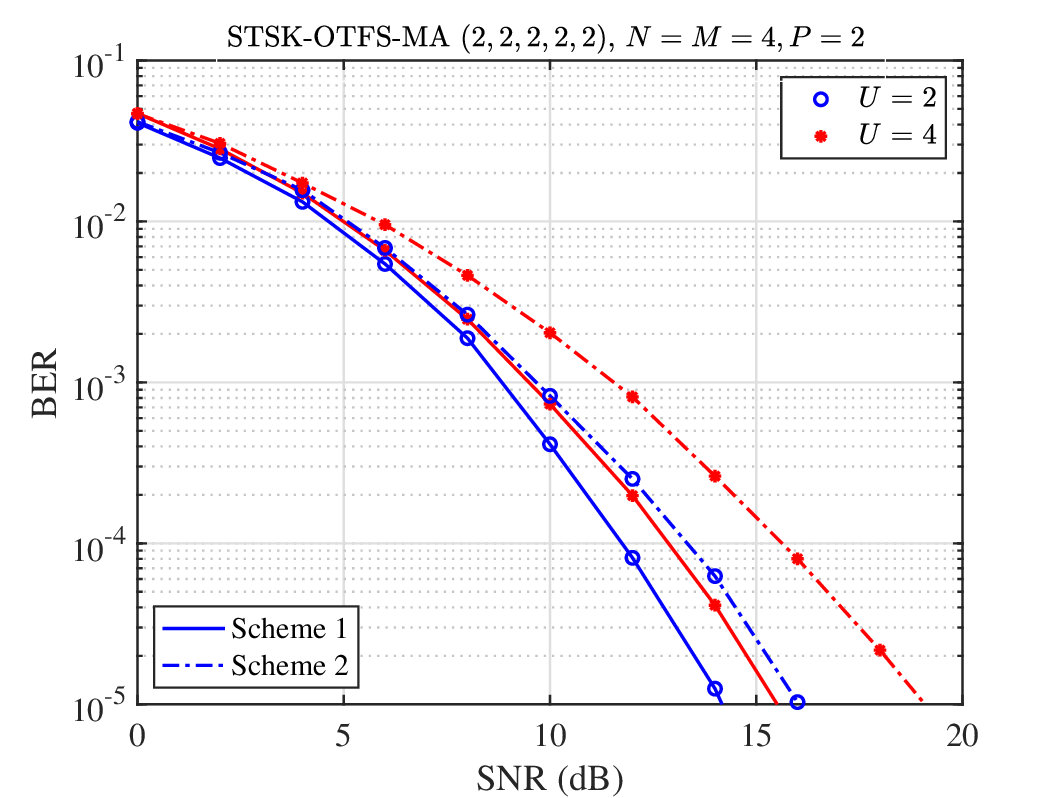}
\caption{BER performance of the STSK-OTFS-MA $(2,2,2,2,2)$ systems with $N=M=4$ and supporting a different number of users by invoking proposed resource allocation schemes as shown in Fig. \ref{Figure2}.}
\label{Figure6-1}
\end{figure}

{Fig. \ref{Figure6-1} evaluates the BER performance of STSK-OTFS-MA $(2,2,2,2,2)$ systems supporting $U=2$ and $U=4$ users by exploiting the proposed resource allocation schemes shown in Fig. \ref{Figure2}. Observe from Fig. \ref{Figure6-1} that the Scheme 1-based system is capable of attaining better BER performance than the system utilising Scheme 2. Moreover, the performance gap becomes wider when our STSK-OTFS-MA supports more users. Explicitly, at a BER of $10^{-5}$, the two-user Scheme 1-based system attains about $2$ dB SNR gain over its Scheme 2 counterpart, while the gain escalates to $4$ dB, when supporting $U=4$ users. This is because the MUI becomes higher as the number of supported users increases. Additionally, the efficiency of our MUI mitigation and resource allocation Scheme 1 is boldly illustrated.}
\begin{figure}[htbp]
\centering
\includegraphics[width=\linewidth]{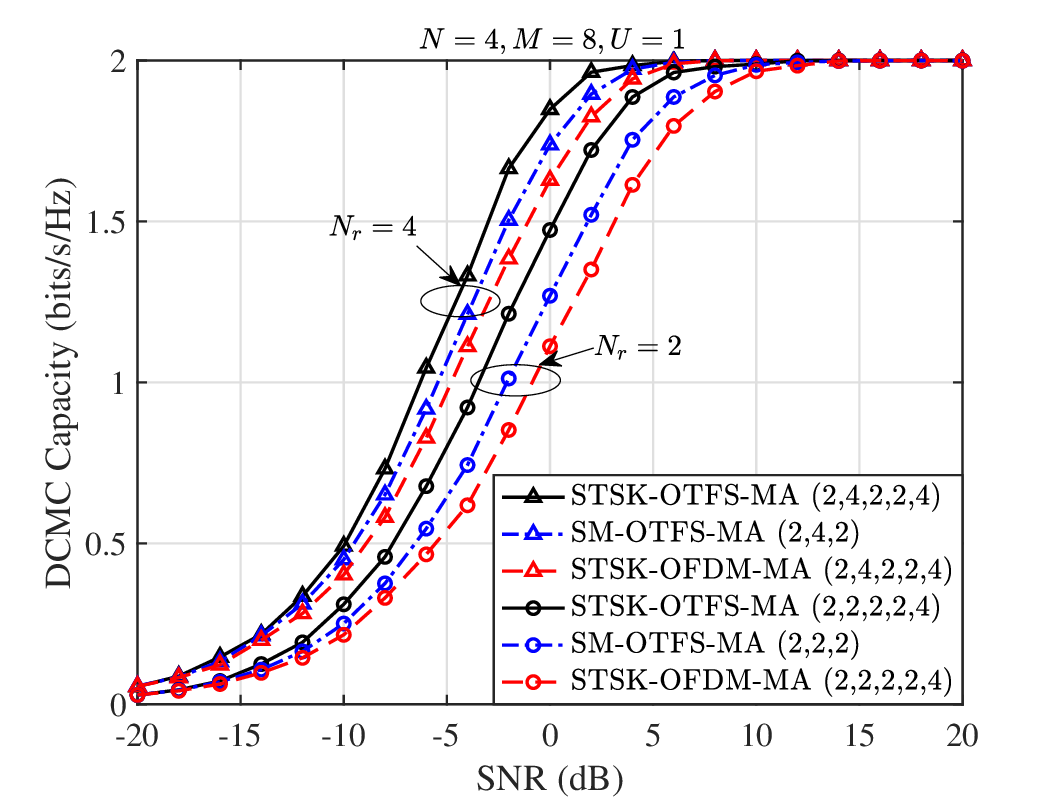}
\caption{The single-user DCMC capacity of the SM-OTFS $(2,N_r,2)$, STSK-OFDM-MA $(2,N_r,2,2,4)$ and our STSK-OTFS-MA $(2,N_r,2,2,4)$ systems with different number of RAs.}
\label{Figure7}
\end{figure}

In Fig. \ref{Figure7}, the single-user DCMC capacities are investigated for both the SM-OTFS $(2,N_r,2)$, {for the STSK-OFDM-MA $(2,N_r,2,2,4)$} and for our STSK-OTFS-MA $(2,N_r,2,2,4)$ schemes at a given rate of $R=2$ bits/s/Hz, where $N_r=2$ and $N_r=4$ are considered respectively. Based on Fig. \ref{Figure7}, we have the following observations. Firstly, the asymptotic capacities of both the SM-OTFS, {STSK-OFDM-MA} and of the STSK-OTFS-MA systems are $R=2$ bits/s/Hz, which is independent of the number of RAs, since the rate was limited to $2$ bits/s/Hz. Furthermore, given a value of $N_r$, it is shown in Fig. \ref{Figure7} that the proposed STSK-OTFS-MA system always outperforms the SM-OTFS and {STSK-OFDM-MA} schemes. This observation can also be inferred from Fig. \ref{Figure6} and \cite{5599264}, since our STSK-OTFS-MA is capable of attaining extra time diversity and ST coding gains over the SM-OTFS scheme, {while the capacity of STSK-OFDM-MA erodes due to the ICI.}
\begin{figure}[htbp]
\centering
\includegraphics[width=\linewidth]{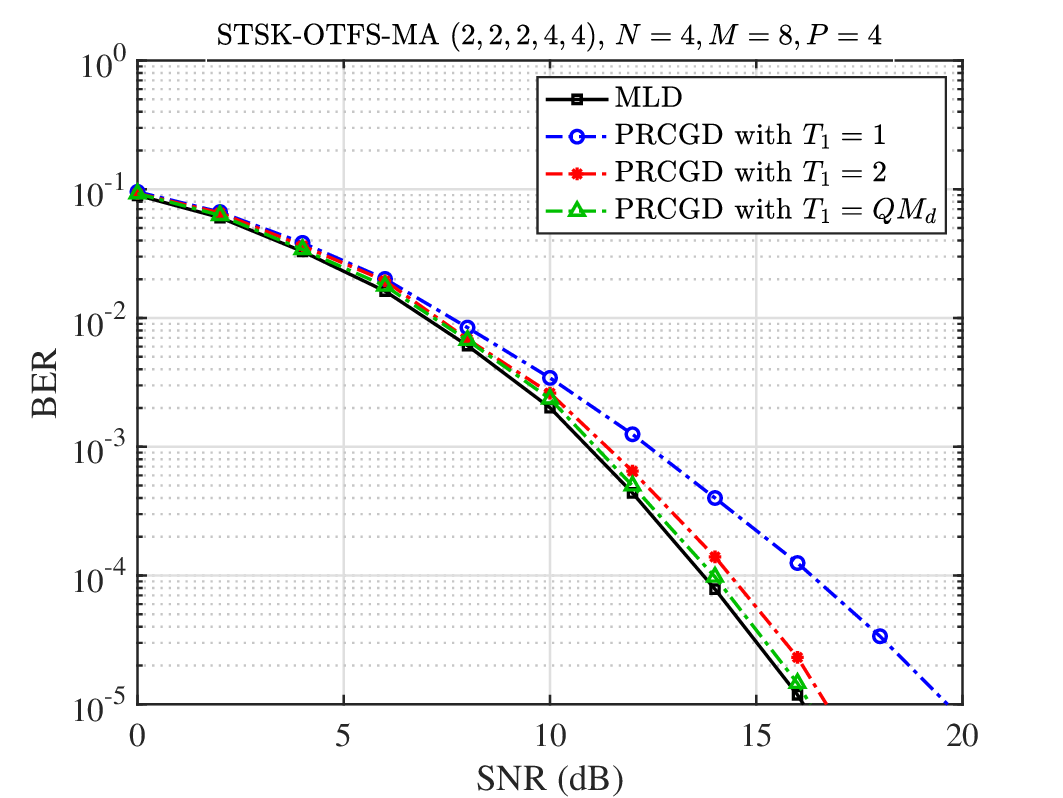}
\caption{Two-user BER performance of the STSK-OTFS-MA $(2,2,2,4,4)$ system using MLD and the proposed PRCGD with the different number of iterations operating at $R=2$ bits/s/Hz.}
\label{Figure8}
\end{figure}

Fig. \ref{Figure8} characterizes the BER performance of both the MLD and PRCGD conceived for the STSK-OTFS-MA $(2,2,2,4,4)$ system operating at $R=2$ bits/s/Hz. We observe from the results of Fig. \ref{Figure8} that the proposed PRCGD relying on two iterations is capable of attaining a BER performance close to that in the $T_1=Q^{M_d}$ case. Moreover, as depicted in Fig. \ref{Figure8}, a near-ML BER performance is attainable when the PRCGD invokes as few as two iterations. It should be noted that $T_1$ denotes the upper-bound of the actual number of iterations in Algorithm \ref{alg1}.

\begin{figure}[htbp]
\centering
\includegraphics[width=\linewidth]{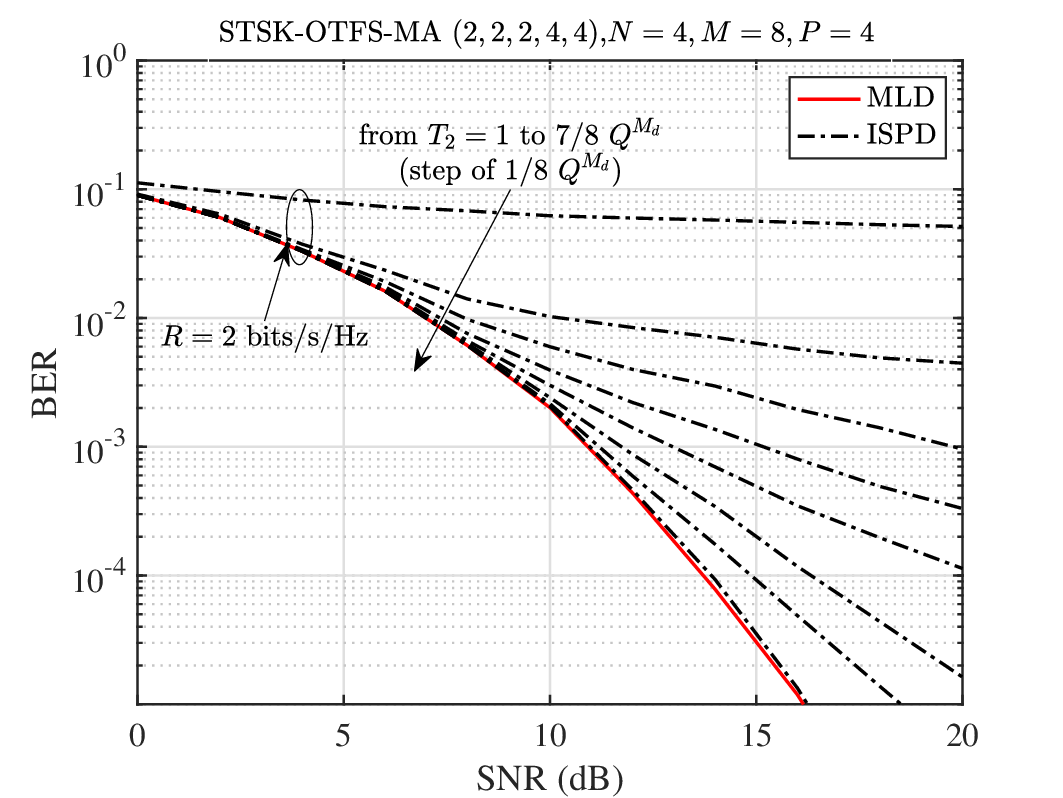}
\caption{Two-user BER performance of the STSK-OTFS-MA $(2,2,2,4,4)$ systems employing MLD and the proposed IRCD with the different number of iterations operating at $R=2$ bits/s/Hz.}
\label{Figure9}
\end{figure}

The BER performance of our IRCD designed for STSK-OTFS-MA is characterized in Fig. \ref{Figure9}, where the BER curve of MLD is depicted as the benchmark, while the other parameters are the same as those for Fig. \ref{Figure8}. It should be emphasized that the IRCD improves a relatively low complexity by invoking an adequate number of $T_2$ than the MLD for the same system. As shown in Fig. \ref{Figure9}, the higher the value of $T_2$, the better BER performance our STSK-OTFS-MA system becomes. Explicitly, in the cases of $T_2>5/8Q^{M_d}$, the IRCD is capable of achieving a better performance than the PRCGD with $T_1=1$, as shown in Fig. \ref{Figure8}. Moreover, we can observe from Fig. \ref{Figure9} that at a BER of $10^{-4}$, the IRCD with $T_2=6/8Q^{M_d}$ attains a gain of about 1.5 dB over the case of using $T_2=5/8Q^{M_d}$, while the IRCD associated with $T_2=7/8Q^{M_d}$ iterations can also achieve a gain of about 1.5 dB over the IRCD with $T_2=5/8Q^{M_d}$. Furthermore, Fig. \ref{Figure9} clearly shows that the IRCD with $T_2=7/8Q^{M_d}$ is capable of attaining nearly the same performance as the MLD. Therefore, from the above observations we conclude that the IRCD with $T_2=5/8Q^{M_d}$ to $T_2=7/8Q^{M_d}$ can be implemented to achieve a desirable BER performance in contrast to the PRCGD associated with $T_1=1$ and $T_1=2$, as shown in Fig. \ref{Figure8}, while improving a considerably lower complexity than the MLD.
\begin{figure}[htbp]
\centering
\includegraphics[width=\linewidth]{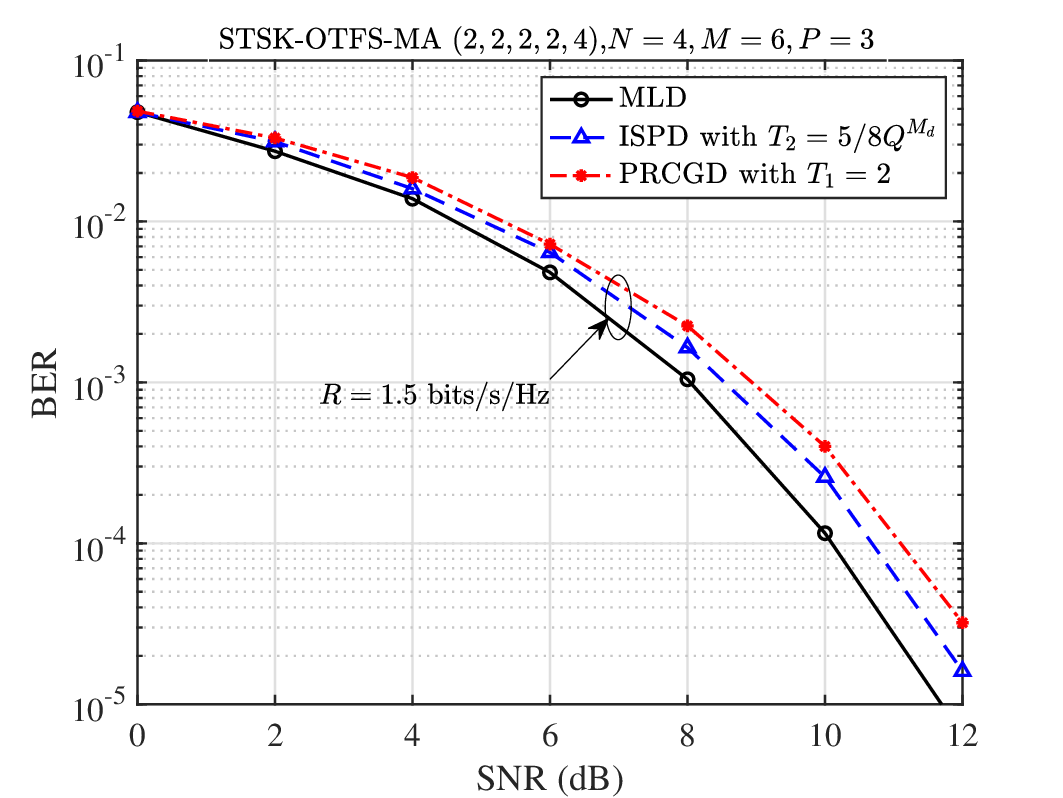}
\caption{Three-user BER performance of the STSK-OTFS-MA $(2,2,2,2,4)$ systems using MLD, our IRCD with $T_1=2$, and the proposed PRCGD with $T_2=5/8Q^{M_d}$ operating at $R=1.5$ bits/s/Hz.}
\label{Figure10}
\end{figure}

To further compare IRCD and PRCGD, in Fig. \ref{Figure10} we characterize the BER performance of these two detectors in three-user STSK-OTFS-MA $(2,2,2,2,4)$ systems, yielding a rate of $R=1.5$ bits/s/Hz. Specifically, the number of iterations is set to $T_1=2$ and $T_2=5/8Q^{M_d}$ for PRCGD and IRCD, respectively. Observe from Fig. \ref{Figure10} that the BER performance of the PRCGD with $T_1=2$ iterations is about 0.5 dB and 1 dB worse than that of the IRCD with $T_2=5/8Q^{M_d}$ and the MLD, respectively. To elaborate further, the PRCGD needs 0.5 dB higher SNR than the IRCD to achieve the BER of $10^{-5}$. Furthermore, it can be concluded that the ISPD associated with $T_2=5/8Q^{M_d}$ iterations obtain a good BER performance compared to the MLD, despite its lower complexity.
\begin{figure}[htbp]
\centering
\includegraphics[width=\linewidth]{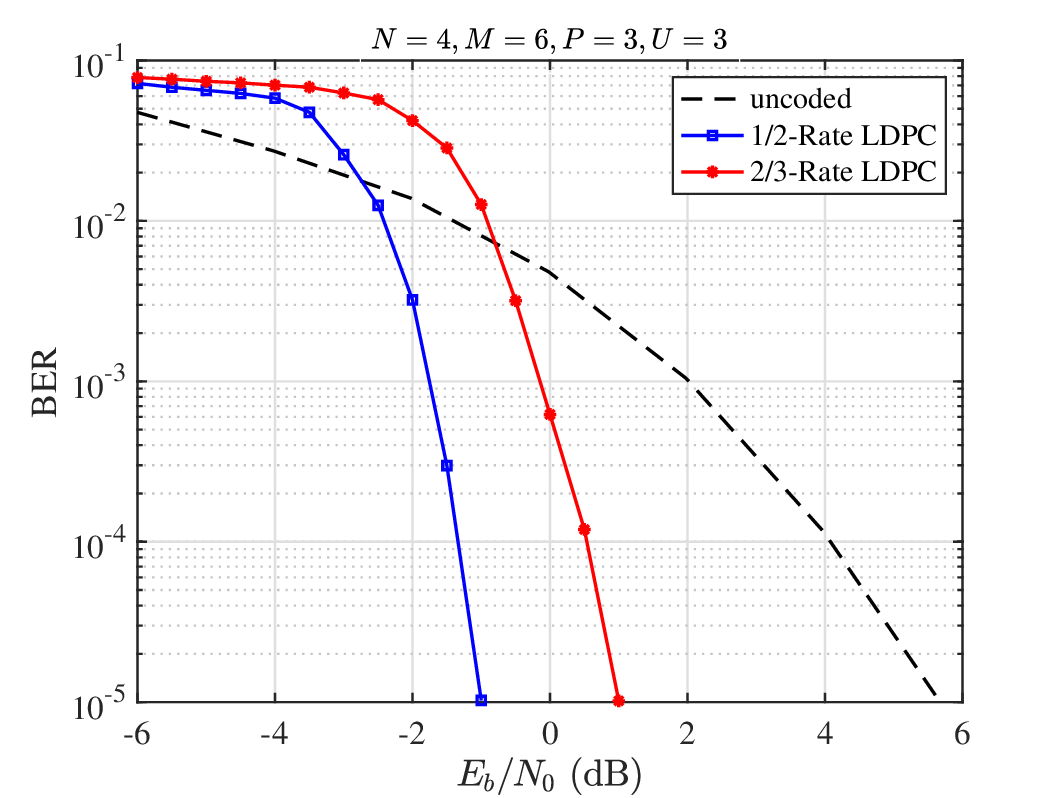}
\caption{BER performance of both the uncoded, rate-$1/2$ and rate-$2/3$ LDPC coded multiuser STSK-OTFS-MA $(2,2,2,2,4)$ systems invoking MLD.}
\label{Figure11-1}
\end{figure}

{To illustrate the general flexibility of our STSK-OTFS-MA scheme, the BER performance of both uncoded, $1/2$-rate and $2/3$-rate LDPC coded STSK-OTFS-MA $(2,2,2,2,4)$ systems using MLD are evaluated in Fig. \ref{Figure11-1}. All the remaining parameters are consistent with those in Fig. \ref{Figure10}. In this context, the sum-product decoding algorithm is harnessed \cite{mackay1997near}. As observed in Fig. \ref{Figure11-1}, the LDPC-coded system is capable of attaining an a substantial performance improvement compared to the conventional uncoded system. Moreover, at a BER of $10^{-5}$, the $1/2$-rate LDPC coded system attains about $2$ dB and $7$ dB SNR gain compared to the $2/3$-rate LDPC coded and uncoded systems, respectively.}
\begin{figure}[htbp]
\centering
\includegraphics[width=\linewidth]{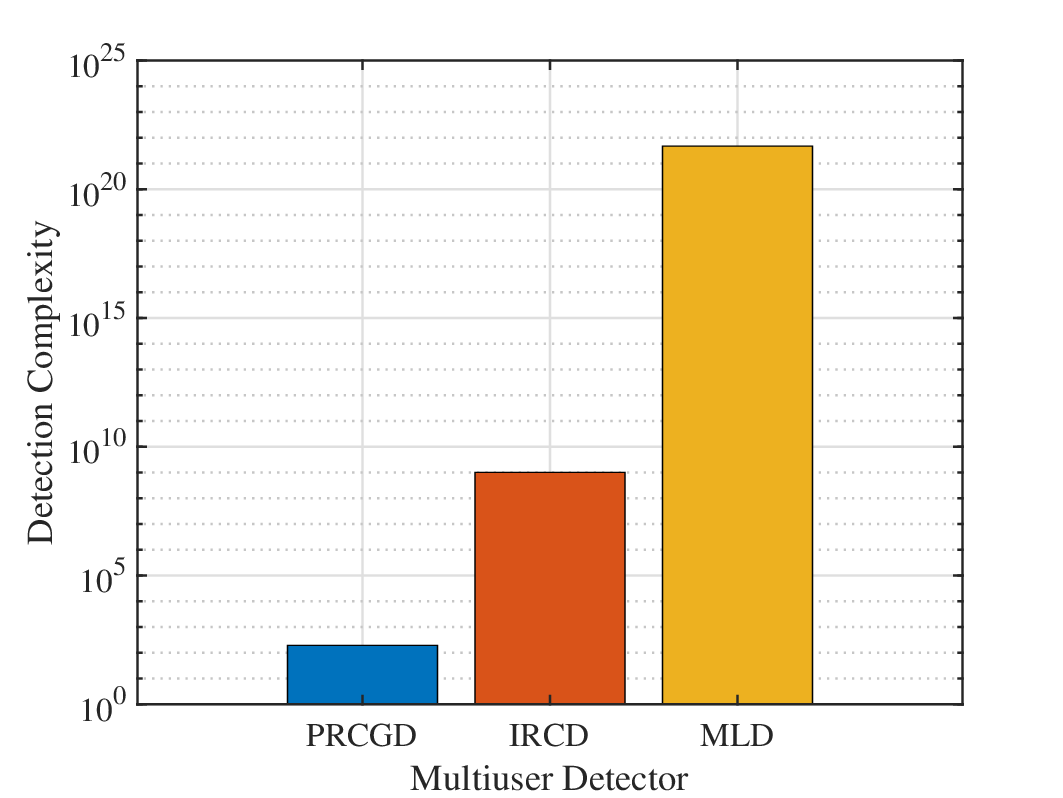}
\caption{Multiuser detection complexity of the STSK-OTFS-MA $(2,2,2,2,4)$ systems invoking MLD, our IRCD with $T_1=2$, and the proposed PRCGD with $T_2=5/8Q^{M_d}$ operating at $R=1.5$ bits/s/Hz.}
\label{Figure11}
\vspace{-1em}
\end{figure}

Fig. \ref{Figure11} portrays the corresponding computational complexity of the MLD, IRCD, and PRCGD employed in Fig. \ref{Figure10}. We have the following observations based on Fig. \ref{Figure11}. Firstly, the complexity of PRCGD with $T_1=2$ is much lower than that of IRCD and of MLD. This can be explained by the fact that our PRCGD tests each DAP uniquely, and repeated searches can be avoided, as illustrated in Algorithm \ref{alg1}. Moreover, the PRCGD employs simple symbol-based detection for APM symbols, whereas the MLD detects all the APM symbols jointly, which can be seen by comparing \eqref{Eq34} and \eqref{Eq43}. Secondly, the IRCD with $T_2=5/8Q^{M_d}$ can provide about 12 orders of magnitude complexity reduction over MLD. Since the reliability sorting of all the DAPs is exploited, the full-research process of MLD can be avoided in the proposed IRCD, yielding a near-ML performance at a significantly lower complexity than the MLD. Finally, based on Fig. \ref{Figure10} and Fig. \ref{Figure11}, it can be concluded that satismetricy BER performances can be attained by invoking both the PRCGD with $T_1=2$ and IRCD with $T_2=5/8Q^{M_d}$, which impose much lower complexity than that of the MLD. Furthermore, the PRCGD with $T_1=2$ attains a more attractive BER vs. complexity trade-off than the IRCD.
\begin{figure}[htbp]
\vspace{-1em}
\centering
\includegraphics[width=\linewidth]{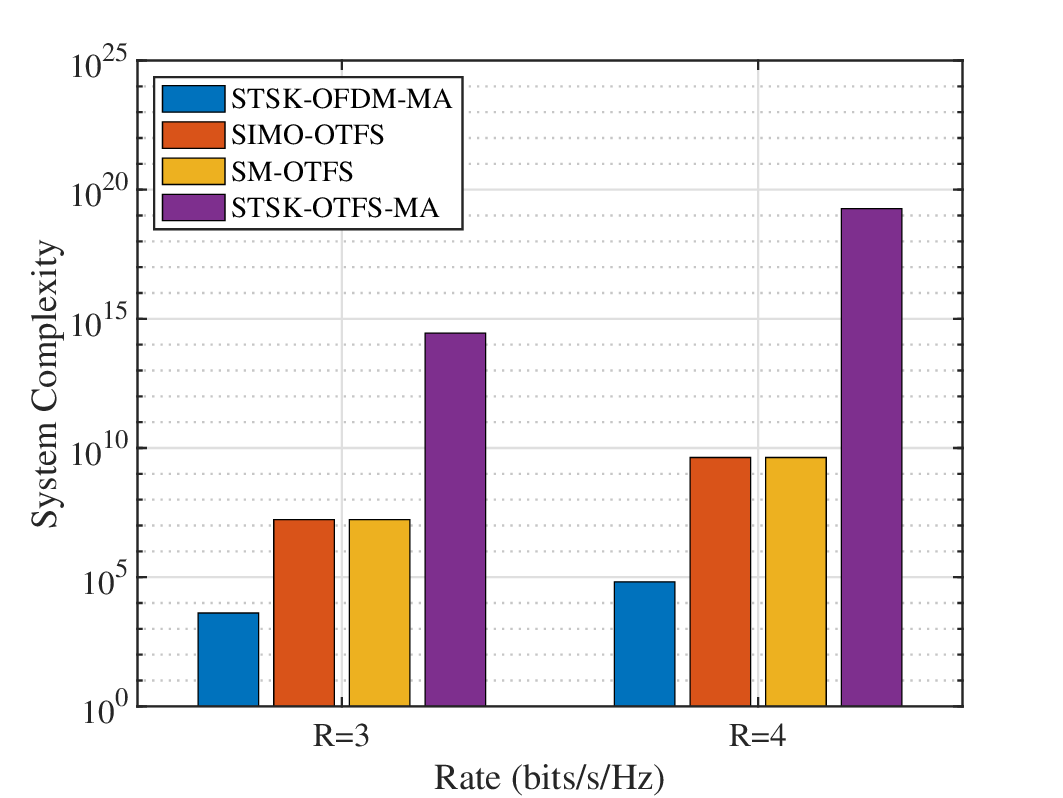}
\caption{System complexity of the conventional SIMO-OTFS scheme, the SM-OTFS scheme, the STSK-OFDM-MA scheme, and our STSK-OTFS-MA scheme invoked in Fig. \ref{Figure6} for a transmission rate of $R=3$ and $R=4$ bits/s/Hz.}
\label{Figure12}
\vspace{-0.5em}
\end{figure}

{In Fig. \ref{Figure12}, the system complexity of the conventional STSK-OFDM-MA, SIMO-OTFS, SM-OTFS and the proposed STSK-OTFS-MA employed in Fig. \ref{Figure6} is investigated. It is observed that STSK-OFDM-MA exhibits the lowest system complexity at a given rate among the OTFS-based systems, followed by the SIMO-OTFS and SM-OTFS paradigms. This is because $N=1$ is invoked in the STSK-OFDM-MA scheme, which significantly reduces the system complexity. However, as shown in Fig. \ref{Figure6}, the BER performance of STSK-OFDM-MA is the worst. The best-performing STSK-OTFS-MA imposes the highest system complexity at a given rate as seen in Fig. \ref{Figure6} and Fig. \ref{Figure7}. Hence, it is demonstrated that our proposed STSK-OTFS-MA strikes a beneficial performance \emph{vs.} system complexity trade-off.}

\section{Summary and Conclusions}\label{Section6}
An STSK-OTFS-MA system has been proposed, where each DD-domain APM symbol is spread over both the space and time dimensions by invoking DMs. Our theoretical derivations illustrated that the proposed STSK-OTFS-MA scheme takes full advantage of both time, frequency, space diversity and also attains ST coding gains. Then, a DD-domain RB allocation scheme has been conceived to mitigate the MUI. Moreover, a pair of low-complexity detectors have been proposed for STSK-OTFS-MA based on greedy algorithms and a codebook of DAPs. Furthermore, based on the MGF technique, the asymptotical BER upper-bound of single-user STSK-OTFS-MA has been derived. Our simulation results have shown that the upper-bound becomes tight at high SNRs. Additionally, the DCMC capacity of our STSK-OTFS-MA scheme has been quantified. Finally, by jointly leveraging the DCMC capacity and the BER union-bound, attractive DM design criteria have been proposed for attaining the maximum attainable diversity and coding gains. Both the analytical and simulation results have demonstrated the superiority of our STSK-OTFS-MA system in terms of both its BER and DCMC capacity. We also demonstrated that there exists an optimal combination of the DM sets and the modulation order. Finally, our simulation results demonstrated that both the proposed PRCGD and IRCD are capable of achieving near-ML BER performances at reduced complexity, {while the proposed STSK-OTFS-MA scheme is capable of attaining better BER performance at an acceptive system complexity compared to other counterparts.}
\renewcommand{\refname}{References}
\mbox{} 
\nocite{*}
\bibliographystyle{IEEEtran}
\bibliography{STSK_OTFS_Zeping.bib}

\begin{thebibliography}{10}

\bibitem{chen2018channel}
Zheng Chen and Emil Bj{\"o}rnson.
\newblock Channel hardening and favorable propagation in cell-free massive
  {MIMO} with stochastic geometry.
\newblock {\em IEEE Transactions on Communications}, 66(11):5205--5219, 2018.

\bibitem{9411900}
Kuntal Deka, Anna Thomas, and Sanjeev Sharma.
\newblock {OTFS-SCMA}: A code-domain {NOMA} approach for orthogonal time
  frequency space modulation.
\newblock {\em IEEE Transactions on Communications}, 69(8):5043--5058, 2021.

\bibitem{8786203}
Zhiguo Ding, Robert Schober, Pingzhi Fan, and H.~V~Poor.
\newblock {OTFS-NOMA}: An efficient approach for exploiting heterogenous user
  mobility profiles.
\newblock {\em IEEE Transactions on Communications}, 67(11):7950--7965, 2019.

\bibitem{elfadil2018trellis}
Husam Elfadil, Mehdi Maleki, Nader Behdad, and Hamid~Reza Bahrami.
\newblock Trellis-coded space-time shift keying.
\newblock {\em IEEE Transactions on Communications}, 66(12):5888--5901, 2018.

\bibitem{7925924}
R.~Hadani, S.~Rakib, M.~Tsatsanis, A.~Monk, A.~J. Goldsmith, A.~F. Molisch, and
  R.~Calderbank.
\newblock Orthogonal time frequency space modulation.
\newblock In {\em 2017 IEEE Wireless Communications and Networking Conference
  (WCNC)}, pages 1--6, 2017.

\bibitem{7448828}
Ibrahim~A. Hemadeh, Mohammed El-Hajjar, Seunghwan Won, and Lajos Hanzo.
\newblock Layered multi-group steered space-time shift-keying for
  millimeter-wave communications.
\newblock {\em IEEE Access}, 4:3708--3718, 2016.

\bibitem{6316188}
Mohammad~Ismat Kadir, Shinya Sugiura, Jiayi Zhang, Sheng Chen, and Lajos Hanzo.
\newblock {OFDMA/SC-FDMA} aided space–time shift keying for dispersive
  multiuser scenarios.
\newblock {\em IEEE Transactions on Vehicular Technology}, 62(1):408--414,
  2013.

\bibitem{9860075}
Venkatesh Khammammetti and Saif~Khan Mohammed.
\newblock Spectral efficiency of {OTFS} based orthogonal multiple access with
  rectangular pulses.
\newblock {\em IEEE Transactions on Vehicular Technology}, pages 1--16, 2022.

\bibitem{9864300}
Muye Li, Shun Zhang, Yao Ge, Feifei Gao, and Pingzhi Fan.
\newblock Joint channel estimation and data detection for hybrid {RIS} aided
  millimeter wave {OTFS} systems.
\newblock {\em IEEE Transactions on Communications}, 70(10):6832--6848, 2022.

\bibitem{9404861}
Shuangyang Li, Jinhong Yuan, Weijie Yuan, Zhiqiang Wei, Baoming Bai, and
  Derrick Wing~Kwan Ng.
\newblock Performance analysis of coded {OTFS} systems over high-mobility
  channels.
\newblock {\em IEEE Transactions on Wireless Communications}, 20(9):6033--6048,
  2021.

\bibitem{liu2021message}
Fei Liu, Zhengdao Yuan, Qinghua Guo, Zhongyong Wang, and Peng Sun.
\newblock Message passing-based structured sparse signal recovery for
  estimation of {OTFS} channels with fractional doppler shifts.
\newblock {\em IEEE Transactions on Wireless Communications},
  20(12):7773--7785, 2021.

\bibitem{8322306}
Siyao Lu, Ibrahim~A. Hemadeh, Mohammed El-Hajjar, and Lajos Hanzo.
\newblock Compressed-sensing-aided space-time frequency index modulation.
\newblock {\em IEEE Transactions on Vehicular Technology}, 67(7):6259--6271,
  2018.

\bibitem{9496190}
Yiyan Ma, Guoyu Ma, Ning Wang, Zhangdui Zhong, and Bo~Ai.
\newblock {OTFS-TSMA} for massive internet of things in high-speed railway.
\newblock {\em IEEE Transactions on Wireless Communications}, 21(1):519--531,
  2022.

\bibitem{mackay1997near}
David~JC MacKay and Radford~M Neal.
\newblock Near shannon limit performance of low density parity check codes.
\newblock {\em Electronics letters}, 33(6):457--458, 1997.

\bibitem{4382913}
Raed~Y. Mesleh, Harald Haas, Sinan Sinanovic, Chang~Wook Ahn, and Sangboh Yun.
\newblock Spatial modulation.
\newblock {\em IEEE Transactions on Vehicular Technology}, 57(4):2228--2241,
  2008.

\bibitem{1608632}
Soon~Xin Ng and L.~Hanzo.
\newblock On the {MIMO} channel capacity of multidimensional signal sets.
\newblock {\em IEEE Transactions on Vehicular Technology}, 55(2):528--536,
  2006.

\bibitem{8424569}
P.~Raviteja, Khoa~T. Phan, Yi~Hong, and Emanuele Viterbo.
\newblock Interference cancellation and iterative detection for orthogonal time
  frequency space modulation.
\newblock {\em IEEE Transactions on Wireless Communications},
  17(10):6501--6515, 2018.

\bibitem{9689960}
Jia Shi, Junfan Hu, Yang Yue, Xuan Xue, Wei Liang, and Zan Li.
\newblock Outage probability for {OTFS} based downlink {LEO} satellite
  communication.
\newblock {\em IEEE Transactions on Vehicular Technology}, 71(3):3355--3360,
  2022.

\bibitem{10043628}
Jia Shi, Zan Li, Junfan Hu, Zhuangzhuang Tie, Shuangyang Li, Wei Liang, and
  Zhiguo Ding.
\newblock {OTFS} enabled {LEO} satellite communications: A promising solution
  to severe doppler effects.
\newblock {\em IEEE Network}, pages 1--7, 2023.

\bibitem{5599264}
Shinya Sugiura, Sheng Chen, and Lajos Hanzo.
\newblock Coherent and differential space-time shift keying: A dispersion
  matrix approach.
\newblock {\em IEEE Transactions on Communications}, 58(11):3219--3230, 2010.

\bibitem{9507331}
Zeping Sui, Shefeng Yan, Hongming Zhang, Lie-Liang Yang, and Lajos Hanzo.
\newblock Approximate message passing algorithms for low complexity {OFDM-IM}
  detection.
\newblock {\em IEEE Transactions on Vehicular Technology}, 70(9):9607--9612,
  2021.

\bibitem{10129061}
Zeping Sui, Hongming Zhang, Yu~Xin, Tong Bao, Lie-Liang Yang, and Lajos Hanzo.
\newblock Low complexity detection of spatial modulation aided {OTFS} in
  doubly-selective channels.
\newblock {\em IEEE Transactions on Vehicular Technology}, pages 1--6, 2023.

\bibitem{8686339}
G.~D. Surabhi, Rose~Mary Augustine, and A.~Chockalingam.
\newblock On the diversity of uncoded {OTFS} modulation in doubly-dispersive
  channels.
\newblock {\em IEEE Transactions on Wireless Communications}, 18(6):3049--3063,
  2019.

\bibitem{979326}
G.~Taricco and E.~Biglieri.
\newblock Exact pairwise error probability of space-time codes.
\newblock {\em IEEE Transactions on Information Theory}, 48(2):510--513, 2002.

\bibitem{9794710}
Anna Thomas, Kuntal Deka, Patchava Raviteja, and Sanjeev Sharma.
\newblock Convolutional sparse coding based channel estimation for {OTFS-SCMA}
  in uplink.
\newblock {\em IEEE Transactions on Communications}, 70(8):5241--5257, 2022.

\bibitem{8515088}
Venkatesh \vspace{0mm}Khammammetti and Saif~Khan Mohammed.
\newblock {OTFS}-based multiple-access in high {Doppler} and delay spread
  wireless channels.
\newblock {\em IEEE Wireless Communications Letters}, 8(2):528--531, 2019.

\bibitem{9508932}
Zhiqiang Wei, Weijie Yuan, Shuangyang Li, Jinhong Yuan, Ganesh Bharatula, Ronny
  Hadani, and Lajos Hanzo.
\newblock Orthogonal time-frequency space modulation: A promising
  next-generation waveform.
\newblock {\em IEEE Wireless Communications}, 28(4):136--144, 2021.

\bibitem{10022044}
Haifeng Wen, Weijie Yuan, Zilong Liu, and Shuangyang Li.
\newblock {OTFS-SCMA}: A downlink {NOMA} scheme for massive connectivity in
  high mobility channels.
\newblock {\em IEEE Transactions on Wireless Communications}, pages 1--1, 2023.

\bibitem{yang2009multicarrier}
Lie-Liang Yang.
\newblock {\em Multicarrier communications}.
\newblock John Wiley \& Sons, 2009.

\bibitem{9082873}
Weijie Yuan, Zhiqiang Wei, Jinhong Yuan, and Derrick Wing~Kwan Ng.
\newblock A simple variational bayes detector for orthogonal time frequency
  space ({OTFS}) modulation.
\newblock {\em IEEE Transactions on Vehicular Technology}, 69(7):7976--7980,
  2020.

\bibitem{9492800}
Zhengdao Yuan, Fei Liu, Weijie Yuan, Qinghua Guo, Zhongyong Wang, and Jinhong
  Yuan.
\newblock Iterative detection for orthogonal time frequency space modulation
  with unitary approximate message passing.
\newblock {\em IEEE Transactions on Wireless Communications}, 21(2):714--725,
  2022.

\bibitem{8478829}
Hongming Zhang, Chunxiao Jiang, Lie-Liang Yang, Ertugrul Basar, and Lajos
  Hanzo.
\newblock Linear precoded index modulation.
\newblock {\em IEEE Transactions on Communications}, 67(1):350--363, 2019.

\end{thebibliography}

\end{document}